\documentclass[
preprint,
amsmath,amssymb,
aps,
prb,
]{revtex4-2}
\usepackage{graphicx}
\usepackage{dcolumn}
\usepackage{bm}
\usepackage{subcaption}
\usepackage{tabularx}
\usepackage{amsmath}
\usepackage{mathtools}
\usepackage{xcolor}
\usepackage{hyperref}
\usepackage{multirow}
\usepackage{tikz}
\usepackage{array}
\usepackage[normalem]{ulem}
\usepackage{booktabs}
\usepackage{array}
\usepackage{amsmath}
\usepackage{amssymb}
\usepackage{verbatim}

\usepackage{color, colortbl}
\usepackage{hhline}
\usepackage[maxfloats=256]{morefloats}
\definecolor{Gray}{gray}{0.9}
\newcommand{\ket}[1]{|#1\rangle}
\newcommand{\bra}[1]{\langle #1 |}
\begin{document}

\title{Quantum phase transition in transverse-field Ising model \\ on Sierpiński gasket lattice }

\author{Tymoteusz Braciszewski}
\email{e-mail: tymbra@st.amu.edu.pl}
    \affiliation{Faculty of Physics and Astronomy, Adam Mickiewicz University,\\
Uniwersytetu Poznańskiego 2, 61-614 Pozna\'n, Poland}

\author{Oliwier Urbański}
\email{e-mail: oliurb4@st.amu.edu.pl}
     \affiliation{Faculty of Physics and Astronomy, Adam Mickiewicz University,\\
Uniwersytetu Poznańskiego 2, 61-614 Pozna\'n, Poland}

\author{Piotr Tomczak}
\email{e-mail: ptomczak@amu.edu.pl}
	\affiliation{Faculty of Physics and Astronomy, Adam Mickiewicz University,\\
Uniwersytetu Poznańskiego 2, 61-614 Pozna\'n, Poland}

\date{\today}
\begin{abstract}
We investigate the quantum phase transition in the transverse-field Ising model on the Sierpiński gasket using finite-size scaling (FSS) and numerical renormalization group (NRG). Since next generations of the fractal lattice contain exponentially more spins, which in turn increase exponentially the Hilbert space dimension, we challenge and prove usefulness of small systems in FSS. We identified a quantum critical point at
  $\lambda_c \approx 2.63 - 2.93$, with critical exponents
  $ \nu \approx 0.64 - 0.71, \beta \approx 0.30, \gamma \approx 1.67$ and $z \approx 1.33$. The numerical renormalization group method produced results consistent with finite-size scaling approach ($\lambda_c = 2.766$$, \beta = 0.316$), supporting our findings. Compared to the values reported so far in the literature, critical field is in a strong disagreement, while exponents are generally similar excluding $\beta$ and $\gamma$. However, it should be noted, that the lattice investigated in previous works is different from ours, while the latter is in our opinion the standard Sierpiński gasket.

\end{abstract}

\maketitle

\section{Introduction}
\label{sec:intro}

Quantum phase transitions in spin systems on fractal lattices with non-integer Hausdorff dimensions provide a valuable framework for investigating the complex interplay between
lattice geometry and the universality of critical phenomena~\cite{Gefen1981,Gefen1984,Tomczak1994}. The transverse-field Ising model (TFIM) is defined on the Sierpiński gasket -- a fractal lattice with Hausdorff dimension of
$d_H = \ln3/\ln2 \approx 1.585$.

The principal motivation for this study stems from the work of Yi \cite{Yi2015} and Krcmar et al. \cite{Krcmar2018}.
Krcmar et al. \cite{Krcmar2018} seem to have utilized the same lattice as Yi \cite{Yi2015},
although their figures indicate otherwise.
These two studies constitute the only investigations of the quantum phase transition
of the TFIM on the Sierpiński gasket fractal lattice,
highlighting the necessity of resolving this ambiguity.
This uncertainty originates from a comparison of Fig. 1 in \cite{Krcmar2018} and Fig. 5(a) in \cite{Yi2015}. The former clearly shows position of the spins (which is rather not standard as for Sierpiński gasket, but an understandable choice for tensor networks). The latter shows just the Sierpiński gasket without precisely showing spins. If the two lattices are identical, the perfect agreement of the results from \cite{Krcmar2018} and \cite{Yi2015} is understandable, but the standard Sierpiński gasket geometry (as the last in Fig.~\ref{SG_lattices}) was not treated in the literature. Otherwise, the coincidence between \cite{Krcmar2018} and \cite{Yi2015} would be peculiar and very unlikely, because changed coordination number from 3 to 4 should at least alter the critical point. Therefore, we assume identical lattices used by the mentioned works and investigate the slightly different, in our opinion the "original", Sierpiński gasket lattice.

Another motivation for the work is the fact that exact-diagonalization is conceptually the most straightforward technique, but in case of fractal lattices faces a doubly exponential computational wall. Indeed, since Hilbert space dimension grows exponentially with the number of spins and this number grows exponentially with next fractal lattice generations, we face a doubly exponential growth of complexity. Additionally, diagonalization needs to be performed for many values of the transverse field, so the task becomes very time-consuming. This situation is limiting unless we use small system sizes and hope that proper scaling reveals useful information about the thermodynamic limit. It is exactly what we do, but it needs verification via an independent method.

In the following sections, we demonstrate, using two complementary methods, finite-size scaling (FSS) and the numerical renormalization group (NRG), how to obtain reliable information about the critical couplings and exponents of the TFIM defined on the Sierpiński gasket lattice by analyzing the properties of small systems far from the thermodynamic limit.

The paper is organized into four sections. In Sec.~\ref{TFIM1d_benchmark},
we apply FSS analysis
to the one-dimensional TFIM,
\begin{subequations}
    \label{Hamiltonian}
    \begin{equation}
        H=-J\sum_{i=1}^N \sigma^{x}_i\sigma_{i+1}^x-h\sum_{i=1}^N\sigma^z_i,
    \end{equation}
    or
    \begin{equation}
        H/J=-\sum_{i=1}^N \sigma^{x}_i\sigma_{i+1}^x-\lambda \sum_{i=1}^N\sigma^z_i,
    \end{equation}
\end{subequations}
the first sum runs over all nearest-neighbor pairs on a given lattice, and $\lambda = h/J$ is the dimensionless transverse field strength.
For the FSS we use small systems with up to $N = 18$
spins with periodic boundary conditions. This approach serves as a strong benchmark
for our numerical method. Even with these small systems,
we recover the exact critical point $\lambda_c = 1$ and the
exponents $z = 1$, $\nu=1$, $\beta=0.125$, and $\gamma=1.75$ with relatively small errors,
which shows that this method can be used even outside
the large-size, asymptotic regime.
In the Sec.~\ref{FSS_Sierpinski}, we use the same FSS technique to study the critical behavior of the TFIM on the Sierpiński
gasket using systems with up to $N=15$ spins.
In Sec.~\ref{1D_TFIM_RG}, we revisit the earlier version of the RG method for quantum systems \cite{Jullien1978} and reproduce
some of the original results
for the Ising spin chain in a transverse field. Finally, in Sec.~\ref{Sierp_Gas_NRG}, we apply the same method to
the TFIM on the Sierpiński gasket lattice and find results that are consistent with those from the FSS approach.
The article ends with a short summary.

\section{Finite-size scaling benchmark on the one-dimensional TFIM}
\label{TFIM1d_benchmark}


We select the one-dimensional transverse-field Ising model, which has an exact solution, to evaluate the FSS procedure under challenging conditions, i.e., for small system sizes of $N = 14, 16, 18$ and exclude larger systems with periodic boundary conditions for two primary reasons.
First, the exact diagonalization (ED) procedure must be repeated for multiple values of the parameter $\lambda$ to determine the scaling functions. For larger systems, the Hilbert space dimension increases exponentially, rendering repeated diagonalizations computationally infeasible within a reasonable time frame.
Second, the primary focus of this study is the transverse field Ising model (TFIM) on the Sierpiński gasket, which low generations consist of a comparable number of spins. Benchmarking on systems of similar size enables assessment of the accuracy and reliability of the FSS procedure in this regime. The results obtained for the 1D TFIM can be compared with the exact analytical solution, providing a quantitative measure of the method's error prior to its application to the Sierpiński gasket, where no such solution is available.
Despite the relatively small system sizes, the benchmark results remain realistic. The FSS procedure accurately reproduces the known critical properties of the 1D TFIM, thereby suggesting the applicability of the method to the Sierpiński gasket.

To perform finite-size scaling, the lowest
energies
$E_0(N, \lambda)$ and $E_1(N, \lambda)$
and their corresponding
eigenstates $|\psi_0(N, \lambda)\rangle$ and $|\psi_1(N, \lambda)\rangle$
are obtained numerically using the exact diagonalization (ED) procedure, specifically we used the Lanczos algorithm.

Subsequently we examine FSS of the
\begin{itemize}
 \item
 Binder cumulant
$U = \frac{3}{2}\Big( 1 - \frac{\langle m^4 \rangle}{3 \langle m^2 \rangle^2} \Big) $,
\item
magnetization
$m(N,\lambda)=\frac{1}{N}\langle \psi_0(N, \lambda)| \mbox{abs} (\sum_{i=1}^N \sigma^z_i) | \psi_0(N, \lambda)\rangle$,
 \item
 gap in the energy spectrum\\
$\Delta(N,\lambda)=E_1(N,\lambda)-E_0(N,\lambda)$,
\item
and magnetic susceptibility
$\chi = \frac{\partial m}{\partial \lambda}$.
\end{itemize}

The standard finite-size scaling ansatz for quantum critical points states that the data
for different system sizes fall onto a single curve for appropriately scaled arguments
and scaled function values near the critical point. We call that curve a scaling function.
Putting $\widetilde{\lambda} =  (\lambda-\lambda_c)/\lambda_c$ we have
\begin{equation}
U(\lambda, N) =  \mathcal{B}\!\left(\widetilde{\lambda}\,  L^{1/\nu} \right),
\label{Binder_scal}
\end{equation}
Here, $\mathcal{B}$ is the scaling function for the Binder cumulant.
Similarly,
\begin{eqnarray}
\Delta(\lambda, N)\, L^{z}  &=& \mathcal{G}\bigl(\widetilde{\lambda}\, L^{1/\nu} \bigr),
\label{gap_scal}
\end{eqnarray}
\begin{equation}
m(\lambda, N)\, L^{\beta/\nu} = \mathcal{M}\!\left( \widetilde{\lambda}\, L^{1/\nu} \right),
\label{m_scal}
\end{equation}
and
\begin{equation}
\chi(\lambda, N)\, L^{-\gamma/\nu} =  \mathcal{X}\!\left(  \widetilde{\lambda}\, L^{1/\nu} \right),
\label{chi_scal}
\end{equation}
where $\mathcal{G}$, $\mathcal{M}$ and $\mathcal{X}$ are appropriate scaling functions and $L \sim N^{1/d_H}$ is the linear size of the system ($d_H = 1$ for a chain and later $d_H = \ln 3/\ln 2$ for the gasket).
\begin{figure}[h!]
    \centering
    \begin{minipage}{0.4\textwidth}
        \centering
        \includegraphics[width=\linewidth]{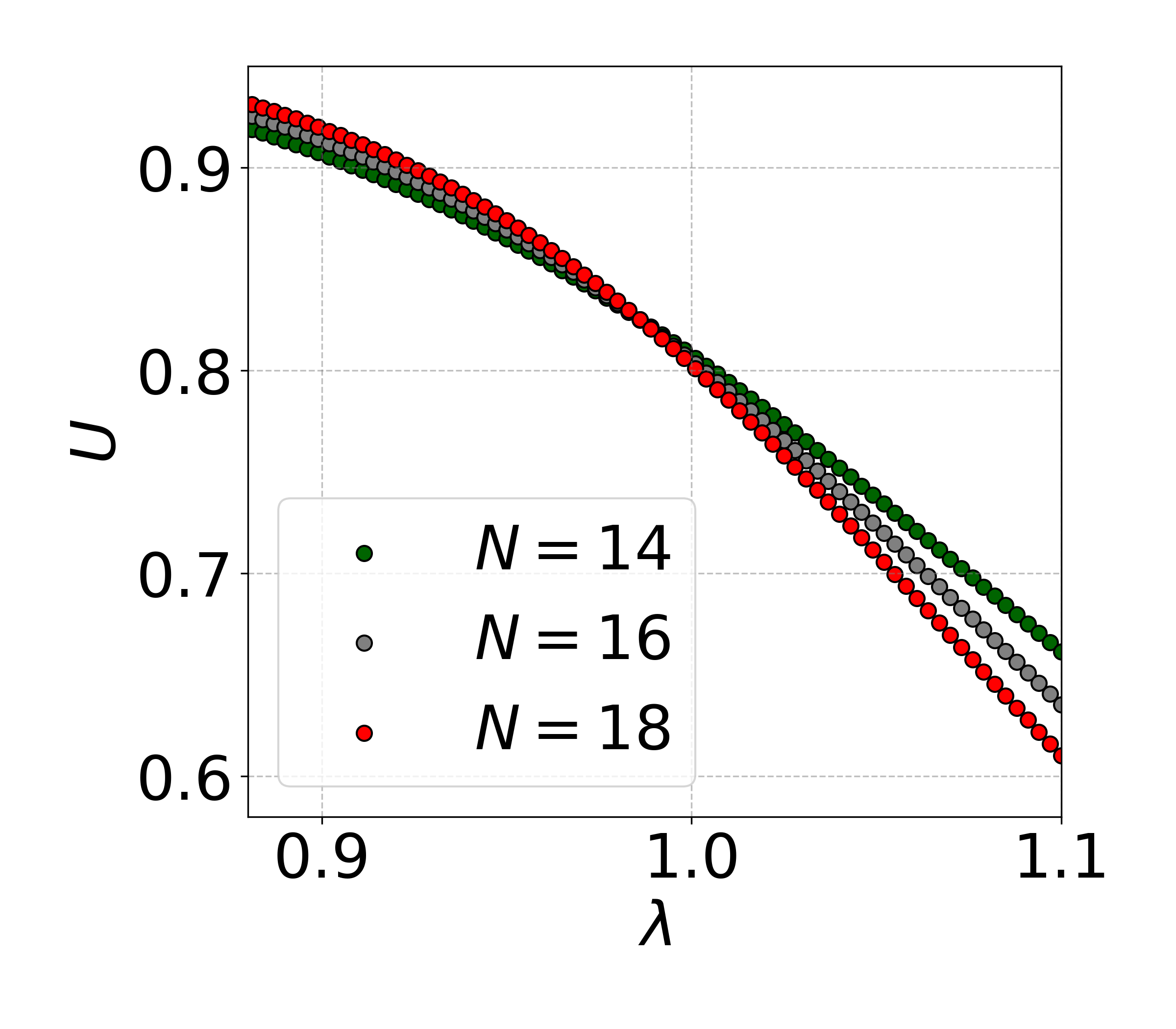}
    \end{minipage}
    \hspace{0.01\textwidth}
    \begin{minipage}{0.4\textwidth}
        \centering
        \includegraphics[width=\linewidth]{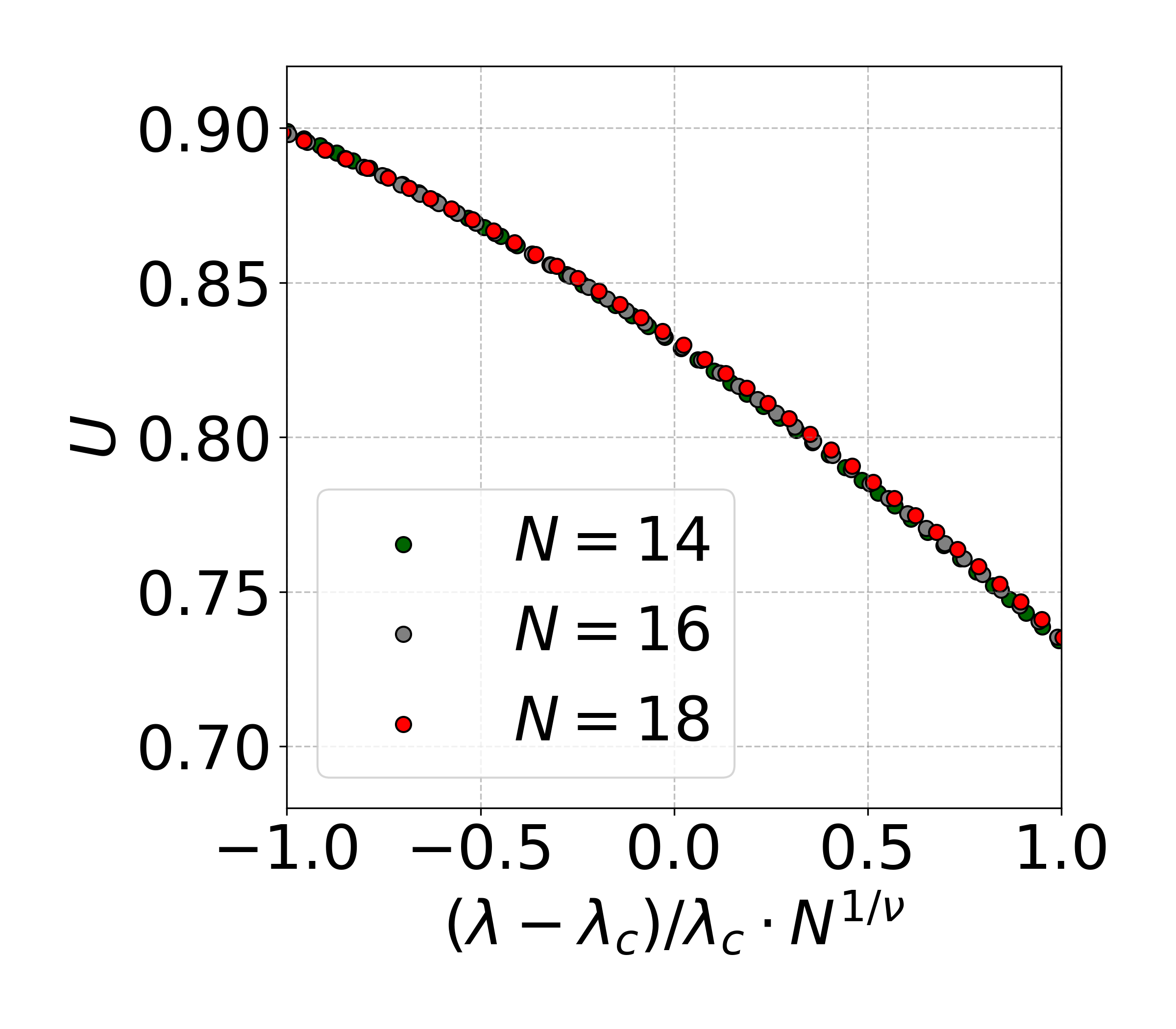}
    \end{minipage}
    \caption{The Binder cumulant before (left) and after
rescaling (right). As expected from the finite size scaling
ansatz, Eq.~(\ref{Binder_scal}), the data for system sizes $N$ = 14, 16, 18 fall
onto a single curve for $ \lambda_c=0.9835, \nu = 1.0044$
(right).}
\label{Binder_scaling_plot}
\end{figure}
Let us examine FSS in detail. Using data from ED as input, we treat $\lambda_c$ and $\nu$ as free fitting parameters
and perform a global nonlinear least-squares fit of Eq.~(\ref{Binder_scal}) to all $U(\lambda, N)$ in the window
$|\widetilde{\lambda} N^\frac{1}{\nu}| < 2$.
Even with only two system sizes $N$ = 18 and 16, the fit converges
and yields:
$ \lambda_c=0.9857, \nu = 1.0050$, see first row and first column of Table \ref{Tr_Is_1D_results}.
The results for scaling systems of other sizes are shown in the first column of this table, while the Binder cumulant
before and after scaling is shown in Fig.~\ref{Binder_scaling_plot}.

The remaining columns of Table \ref{Tr_Is_1D_results} show critical points and exponents from the collapse of magnetization, gap, and susceptibility. FSS slightly underestimates $\lambda_c$ and overestimates $\nu$ when only the exponents related to correlation length divergence $(\nu)$ are used for scaling. This tendency reverses when the exponents $(\beta$ and $\gamma)$ related to the magnetic field
$(\lambda)$ are used together.

\begin{figure}[h!]
    \centering
    \begin{minipage}{0.4\textwidth}
        \centering
        \includegraphics[width=\linewidth]{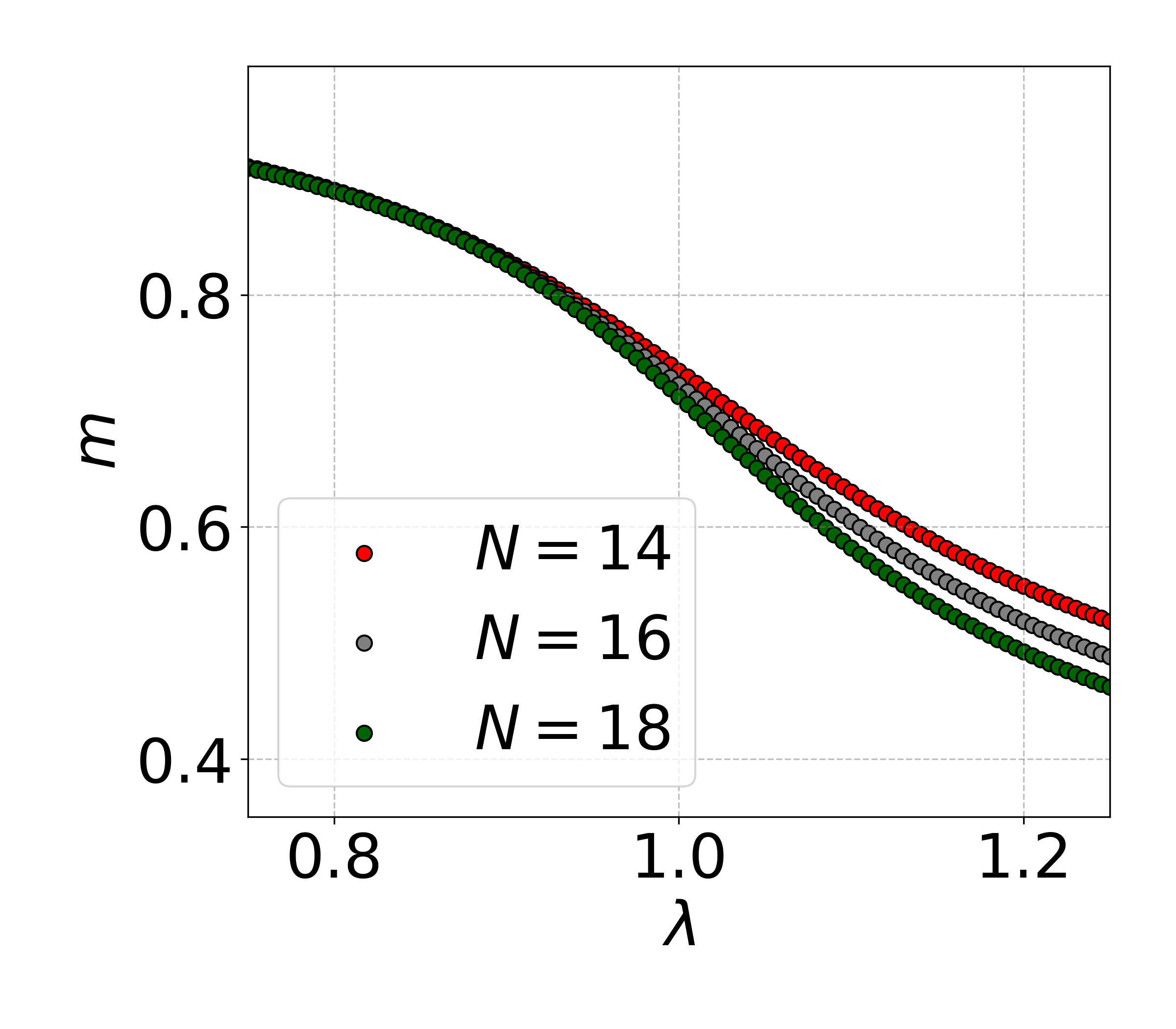}
    \end{minipage}
    \hspace{0.01\textwidth}
    \begin{minipage}{0.4\textwidth}
        \centering
        \includegraphics[width=\linewidth]{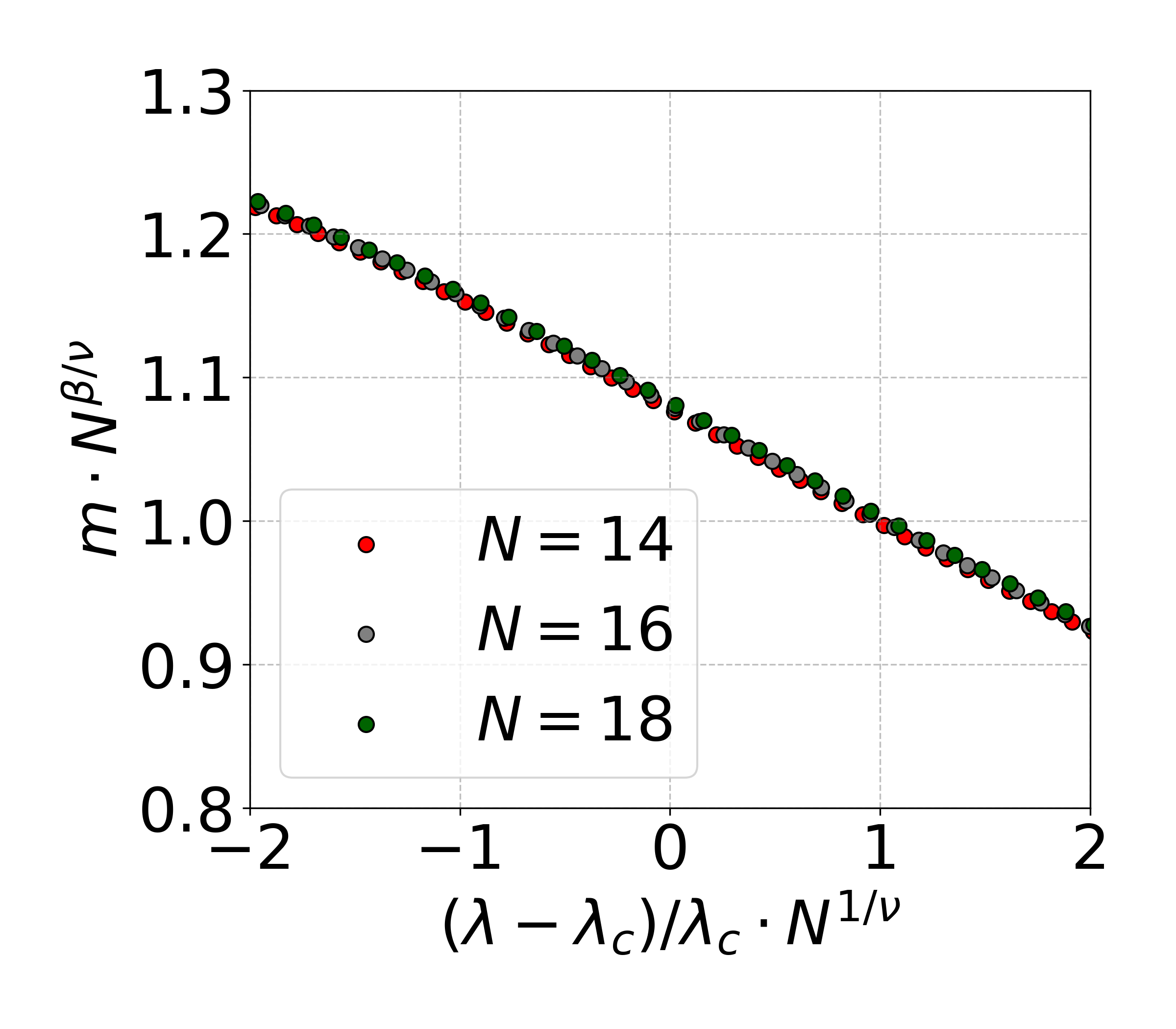}
    \end{minipage}
    \caption{The magnetization before (left) and after rescaling (right)
    according to the Eq.~(\ref{m_scal}). The values of $m$ calculated for system sizes $N=14, 16, 18$ fall
onto a single curve for $\lambda_c=1.0051$, $\beta=0.1253$, and $\nu=0.9157$
(right). This collapse was obtained with fitted parameters differing by  0.3\% from the exact value of $\beta$
and 8\% from the exact value of $\nu$ - it showes the robustness of the FSS procedure even for very small systems.}
\label{magnetization_scaling}
\end{figure}

This behavior is clearly visible in the scaling collapse of the order parameter $m$, see Fig.~\ref{magnetization_scaling}.
Using the system sizes $(N=14-18)$, the optimal
collapse yields $\lambda_c \approx 1.0051$, $\beta \approx 0.1253$, and $\nu \approx 0.9157$.
The value of $\beta$ is already very close to the exact Ising exponent $\beta=1/8$, whereas the estimate of $\nu$ is noticeably smaller than the expected value $\nu=1$.
This discrepancy is attributed to finite-size corrections, which are known to affect slightly scaling collapses.

\begin{figure}[h!]
    \centering
    \begin{minipage}{0.4\textwidth}
        \centering
        \includegraphics[width=\linewidth]{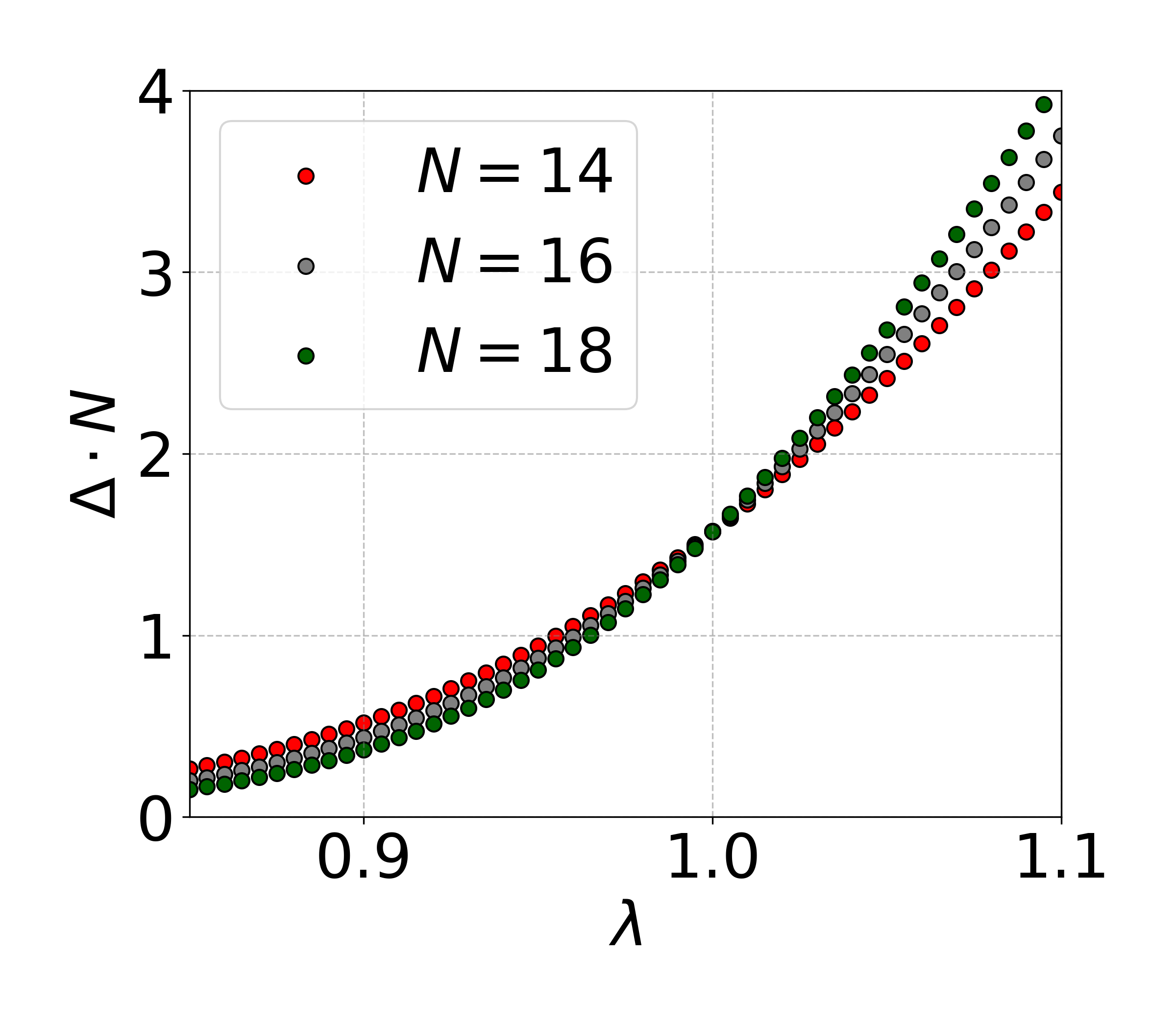}
    \end{minipage}
    \hspace{0.01\textwidth}
    \begin{minipage}{0.4\textwidth}
        \centering
        \includegraphics[width=\linewidth]{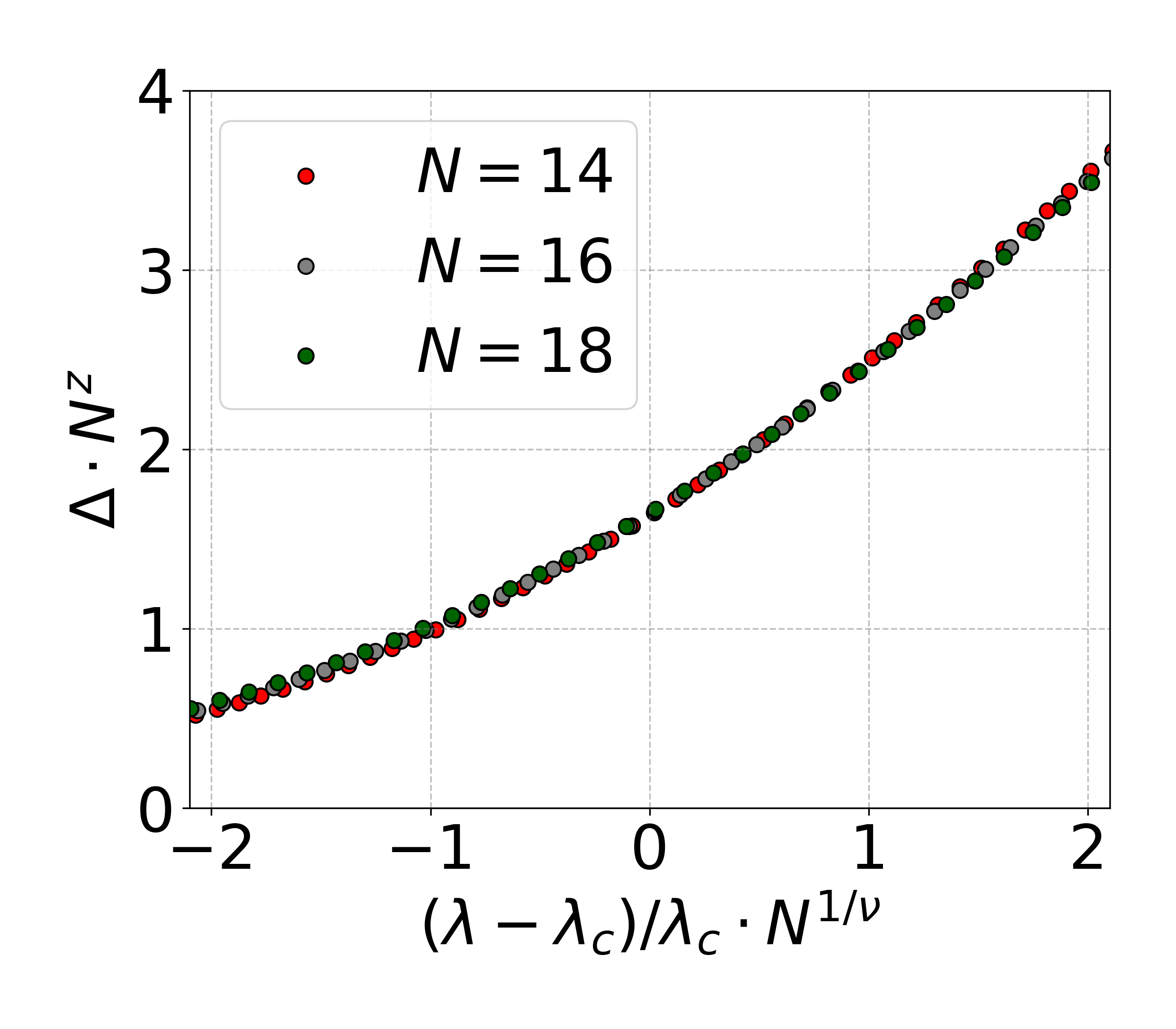}
    \end{minipage}
    \caption{The gap between ground state and first excited state before (left) and after
rescaling (right). The values of the gap calculated for system sizes $N=14, 16, 18$ fall
onto a single curve for $\lambda_c=0.9998$ and $\nu=1.0123$
(right).
To perform this scaling, $z=1$ was assumed. Taking the value of $\nu$ from the scaling, for example of $m$ and fixing it, yields a slightly lower value of $z$; see the text.
}
\label{gap_scaling_plot}
\end{figure}

Let us briefly discuss the gap scaling and the method used to determine the exponent $z$. The fourth column of the Table~\ref{Tr_Is_1D_results}  presents
the values of $\nu$ found by assuming $z=1$.
Taking for the FSS of gap $\Delta(\lambda, N)$
the values of $\nu$ obtained from the scaling of $m$ yields a slightly lower value of $z=0.9648$.
This indicates a minor deviation from the exact value, attributable to the increased sensitivity of $m$
to subleading corrections in small systems.
Figure~\ref{gap_scaling_plot} presents the scaled data collapse. When $\Delta N^{z}$ is plotted against the scaling variable
$\widetilde{\lambda}\, N^{1/\nu}$, the curves for all three system sizes align on a single universal curve with minimal visible scatter.

\begin{table}[h]
\centering
\caption{
Critical points and critical exponents obtained for 1D TFIM from FSS of small systems.
The quantities used for FSS are in the columns, and sizes of systems, critical parameters, and critical exponents
are in the rows.
The last column gives the width of the window in which FSS was performed.
The statistical errors referring to the last signiﬁcant digits are indicated in parentheses.}
\begin{tabular}{lcccccccc}
\toprule
\rowcolor{Gray}
            & $U$& $m$ &  $\Delta$ & $\chi$ & $|\widetilde{\lambda} L^\frac{1}{\nu}|$ \\
\midrule
18 - 16
\\
\midrule
 $\lambda_c$ & 0.9857(28)& 1.0050(7) &  0.9998(3)  &  1.0062(2)& $<\!2.0$ \\
 $\nu$       & 1.0051(20) & 0.9186(47) &  1.0123(20) & 0.8945(53) & $<\!2.0$ \\
 $\beta $    & ---       & 0.1265(13) &  --- & --- & $<\!2.0$ \\
 $\gamma$    & ---       & --- &  --- & 1.7576(18) &$<\!2.0$ \\
\midrule
18 - 16 - 14 \\
\midrule
 $\lambda_c$ & 0.9835(19)& 1.0051(7) & 0.9992 (8)  &  1.0064(6)& $<\!2.0$ \\
 $\nu$       & 1.0044(31) & 0.9157(23) &  1.0360(34) & 0.8926(55) & $<\!2.0$ \\
 $\beta $    & ---       & 0.1253(15) &  --- & --- & $<\!2.0$ \\
 $\gamma$    & ---       & --- &  --- & 1.7529(33) &$<\!2.0$ \\
\midrule
14 - 10 - 6 \\
\midrule
 $\lambda_c$ & 0.9565(36)& 1.0095(31) & 0.9999 (10)  &  1.0226(18)& $<\!1.0$ \\
 $\nu$       & 1.0150(47) & 0.8680(43) &  1.0843(64) & 0.8149(75) & $<\!1.0$ \\
 $\beta $    & ---       & 0.1200(35) &  --- & --- & $<\!1.0$ \\
 $\gamma$    & ---       & --- &  --- & 1.6760(63) &$<\!1.0$ \\
\midrule
14 - 6 \\
\midrule
 $\lambda_c$ & 0.9587(49)& 1.0126(45) & 0.9998 (12)  &  1.0278(29)& $<\!1.0$ \\
 $\nu$       & 1.0143(51) & 0.8673(63) &  1.0806(78) & 0.8340(69) & $<\!1.0$ \\
 $\beta $    & ---       & 0.1236(27) &  --- & --- & $<\!1.0$ \\
 $\gamma$    & ---       & --- &  --- & 1.6135(61) &$<\!1.0$ \\
\bottomrule
\end{tabular}
\label{Tr_Is_1D_results}
\end{table}

The results of the benchmark confirm that our FSS approach, including the choice of fitting window
and numerical differentiation of $m$, is accurate and stable enough to be applied
to the more challenging fractal geometry of the Sierpi\'nski gasket. In this case, exact results
are unavailable, and system sizes must remain modest.

\section{SIERPI\'{N}SKI GASKET - FSS ANALYSIS}
\label{FSS_Sierpinski}

\begin{figure}[h!]
    \centering
   \begin{minipage}{0.25\textwidth}
        \centering
        \includegraphics[width=.75\linewidth]{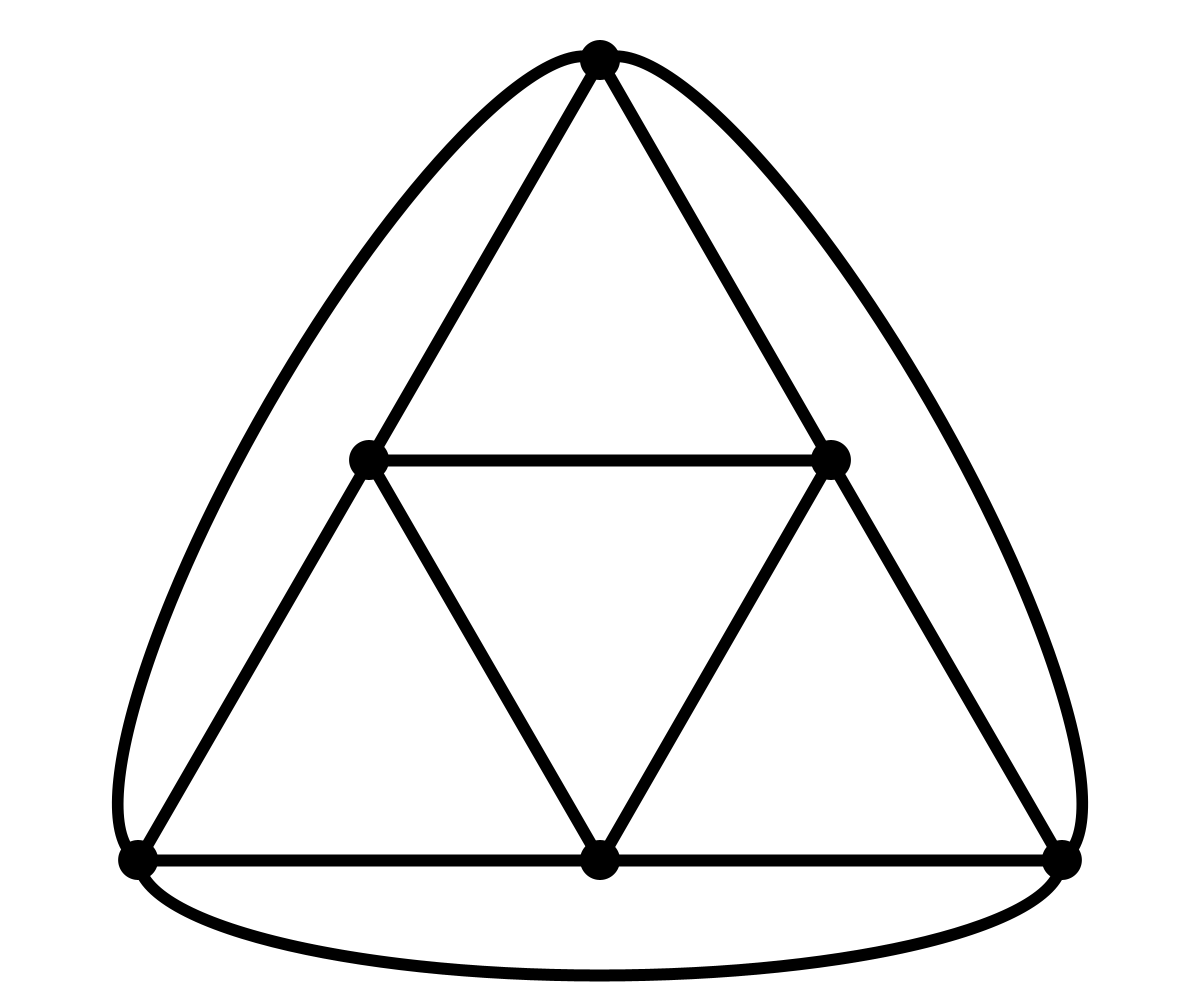}
        \caption*{N=6}
    \end{minipage}
    \begin{minipage}{0.25\textwidth}
        \centering
        \includegraphics[width=\linewidth]{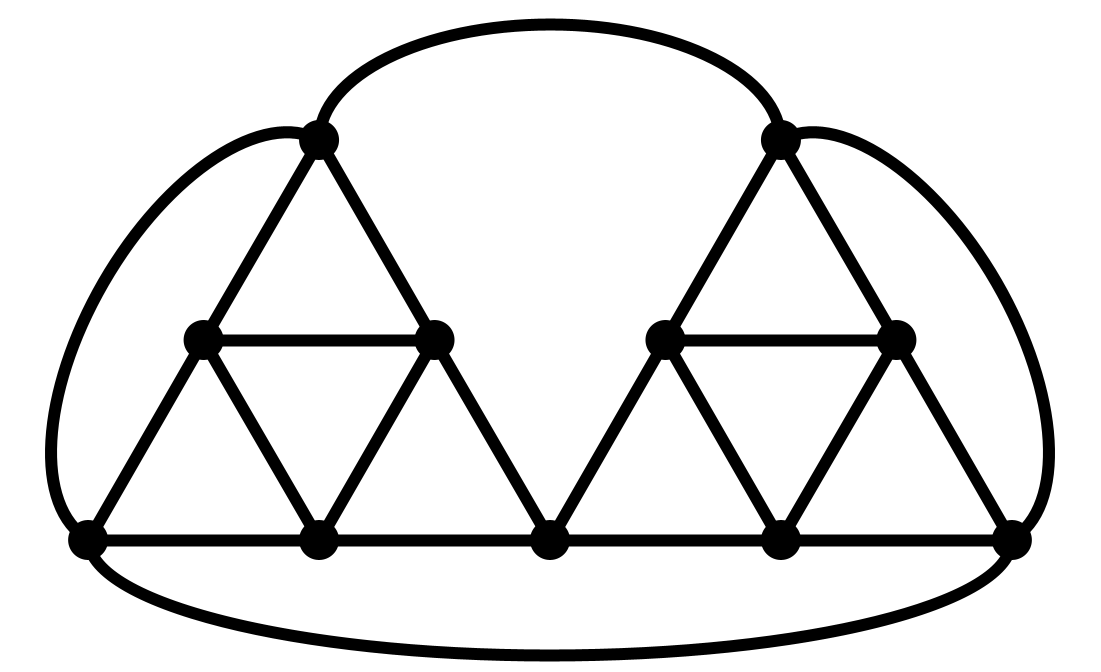}
        \caption*{N=11}
    \end{minipage}
    \hspace{0.01\textwidth}
    \begin{minipage}{0.25\textwidth}
        \centering
        \includegraphics[width=\linewidth]{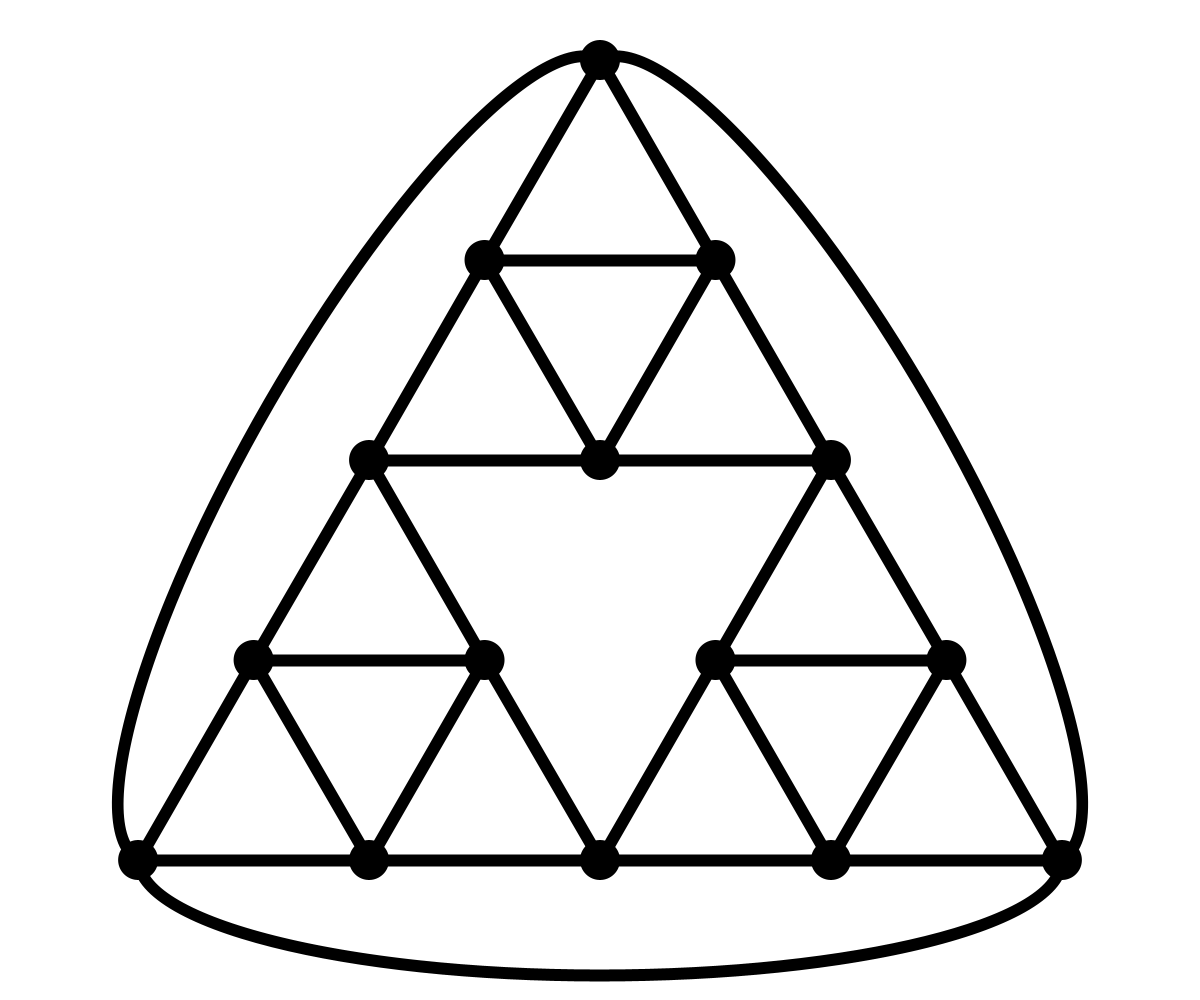}
        \caption*{N=15}
    \end{minipage}
    \caption{Sierpi\'nski gasket lattices with periodic boundary conditions.}
    \label{SG_lattices}
    \end{figure}

Having validated our FSS approach on the exactly solvable one-dimensional chain, we now apply the same methodology to the transverse-field Ising model on the Sierpiński gasket lattice. Like the periodic chain, the gasket has also periodic boundary conditions, i.e., with corner sites having the same coordination number, see Fig.~\ref{SG_lattices}.

We studied gasket systems of sizes $N = 11, 15$ spins, see Fig.~\ref{SG_lattices}. For each system size and field strength $\lambda$, we obtained the ground state $|\psi_0(N,\lambda)\rangle$ and first excited state $|\psi_1(N,\lambda)\rangle$ through exact diagonalization using the Lanczos algorithm. From these eigenstates, we computed the same set of observables as in Section II: the Binder cumulant $U$, magnetization $m$, energy gap $\Delta$, and magnetic susceptibility $\chi$.

Exact diagonalization procedure was enhanced by exploiting symmetries
of the system. Two operators realizing symmetry transformations were
used -- a rotation $R$ by $120^{\circ}$ of the entire lattice and
a global spin flip $F$, with respect to the $x$-axis. Since both of these operators commute with
the Hamiltonian (and with each other), their eigenstates
can be used to block-diagonalize the Hamiltonian.
In other words, energy eigenstates of the system can be chosen to
belong to specific eigenspaces of both $R$ and $F$. For $J,h>0$
mean energy is minimized when quantum amplitudes of all spin configurations
have the same phase. Thus, on the basis of variational principle,
ground state can be chosen to have all amplitudes positive real, so
it necessarily lives in the eigenspace with $R=1$ and $F=1$. Dimensionality
of this subspace is $5472$ for the $15$-spin gasket, which is roughly
equal to $2^{15}/6$. The Hilbert space of the next full stage Sierpiński gasket of $N = 42$ spins has dimension $2^{42} \approx 4.4\times10^{12}$. Exploiting the symmetry reduces this by a factor of 6, but still storage of the three Lanczos vectors required for iterative diagonalization would demand roughly 18 TB of RAM, making it a heavy task. Even though doable, we want to check performance of small systems in FSS.

The finite-size scaling analysis follows the same procedure outlined in Section II, with the scaling ansatz given by Eqs.~(\ref{Binder_scal}--\ref{chi_scal}). We performed global nonlinear least-squares fits treating $\lambda_c$ and the relevant critical exponents as free parameters. The fitting window was restricted to $|\tilde{\lambda}L^{1/\nu}| < 1.0$ to minimize contributions from subleading corrections, which are expected to be more pronounced on the fractal geometry than on the regular chain.

\begin{figure}[h]
\centering
\includegraphics[width=0.45\textwidth]{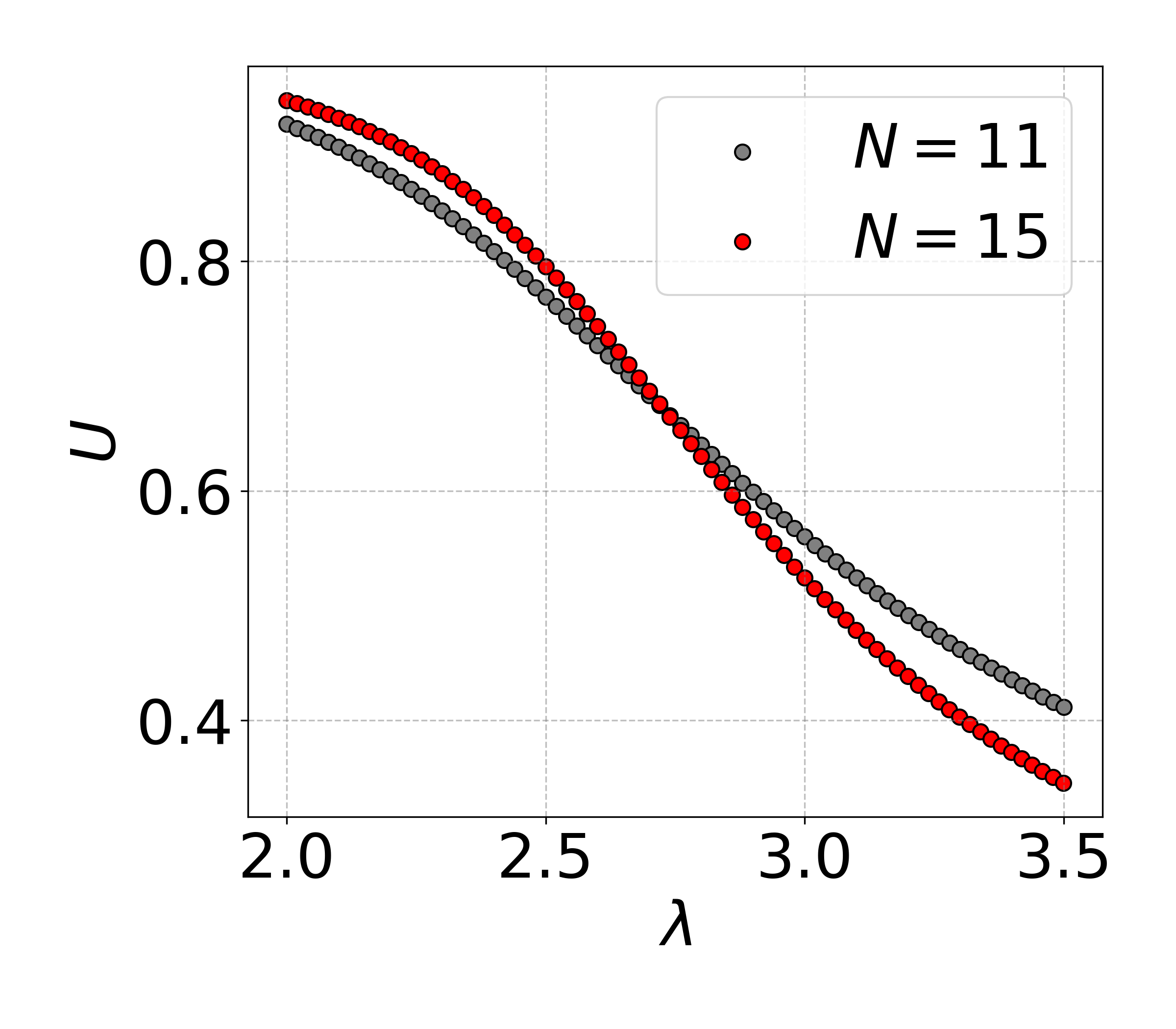}
\includegraphics[width=0.45\textwidth]{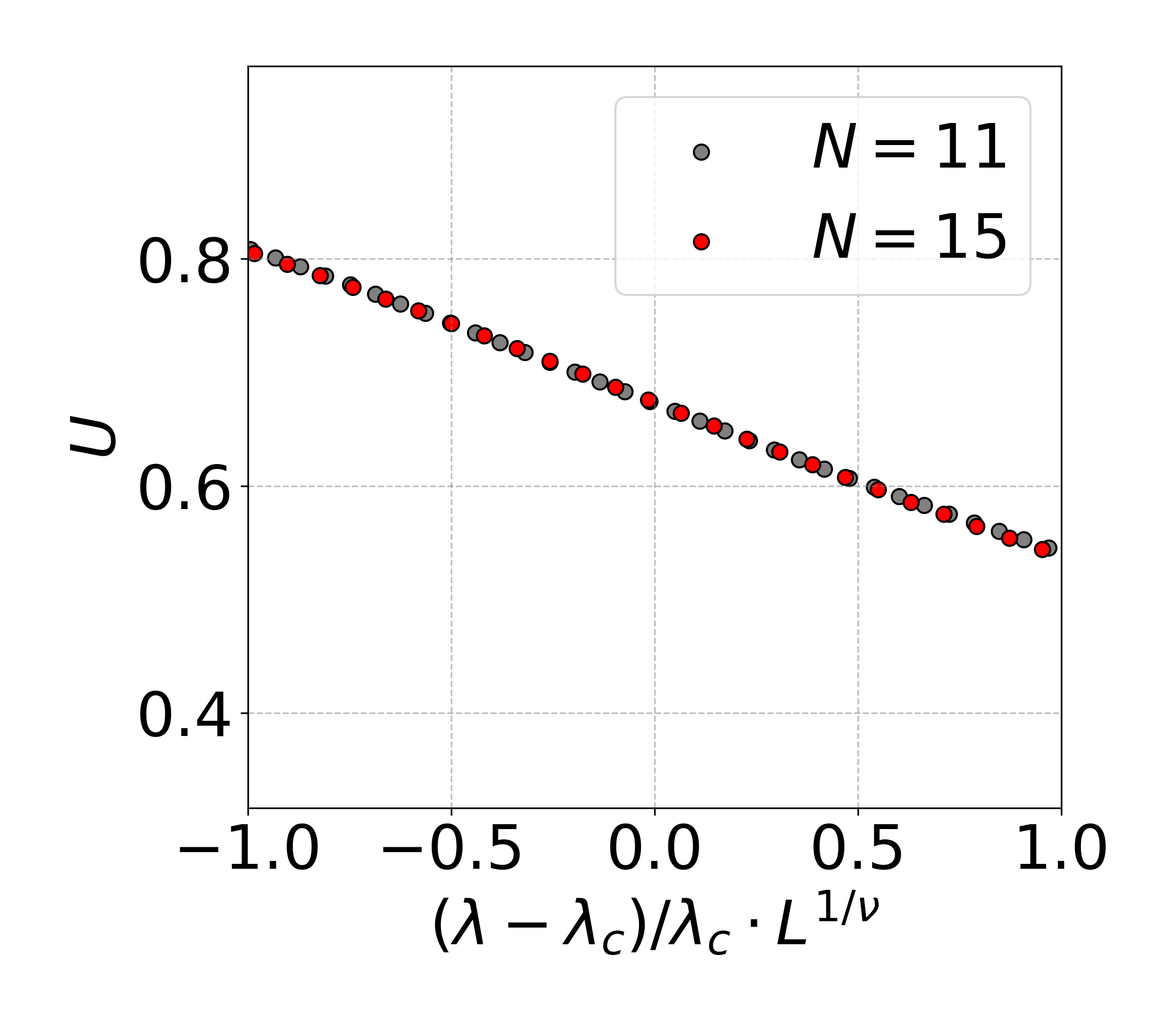}
\caption{The Binder cumulant for the Sierpiński gasket before (left) and after rescaling (right). The data for system sizes $N = 11, 15$ collapse onto a single curve for $\lambda_c = 2.724(2)$ and $\nu = 0.713(1)$.}
\label{fig:gasket_binder}
\end{figure}

Figure~\ref{fig:gasket_binder} shows the Binder cumulant before and after scaling for the two largest system sizes ($N = 11, 15$). The collapse yields $\lambda_c = 2.724(2)$ and $\nu = 0.713(1)$, demonstrating data collapse despite the modest system sizes.

\begin{figure}[h]
\centering
\includegraphics[width=0.45\textwidth]{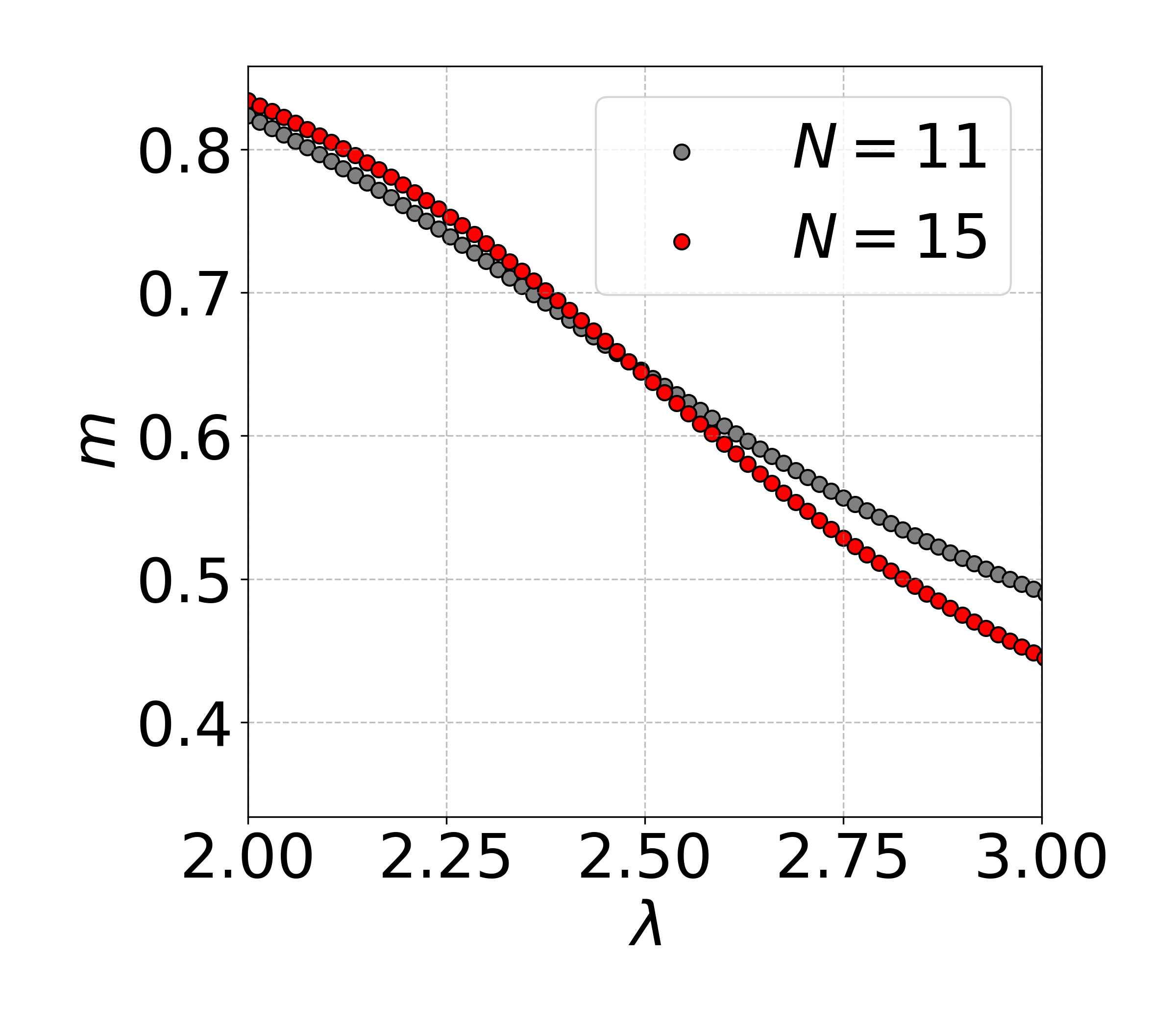}
\includegraphics[width=0.45\textwidth]{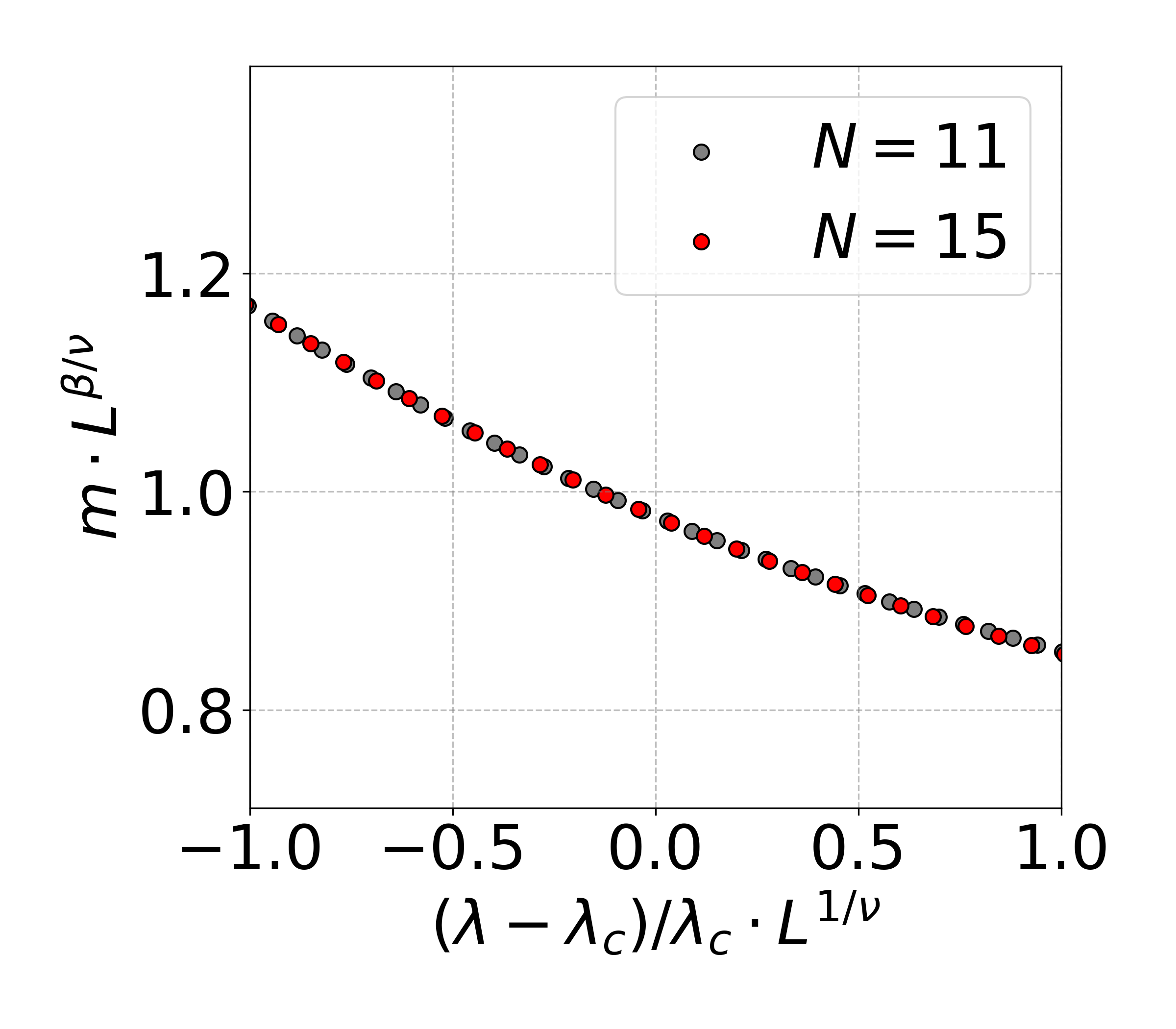}
\caption{The magnetization for the Sierpiński gasket before (left) and after rescaling (right). The collapse for $N = 11, 15$ yields $\lambda_c = 2.930(6)$, $\beta = 0.300(5)$, and $\nu = 0.69(4)$.}
\label{fig:gasket_mag}
\end{figure}

The magnetization scaling, shown in Fig.~\ref{fig:gasket_mag}, yields $\lambda_c = 2.930(6)$, $\beta = 0.300(5)$, and $\nu = 0.69(4)$ for the two-size collapse. The value of $\beta$ is notably larger than the one-dimensional result ($\beta = 1/8$), indicating a strong change in the order parameter behavior due to the fractal geometry.

\begin{figure}[h]
\centering
\includegraphics[width=0.45\textwidth]{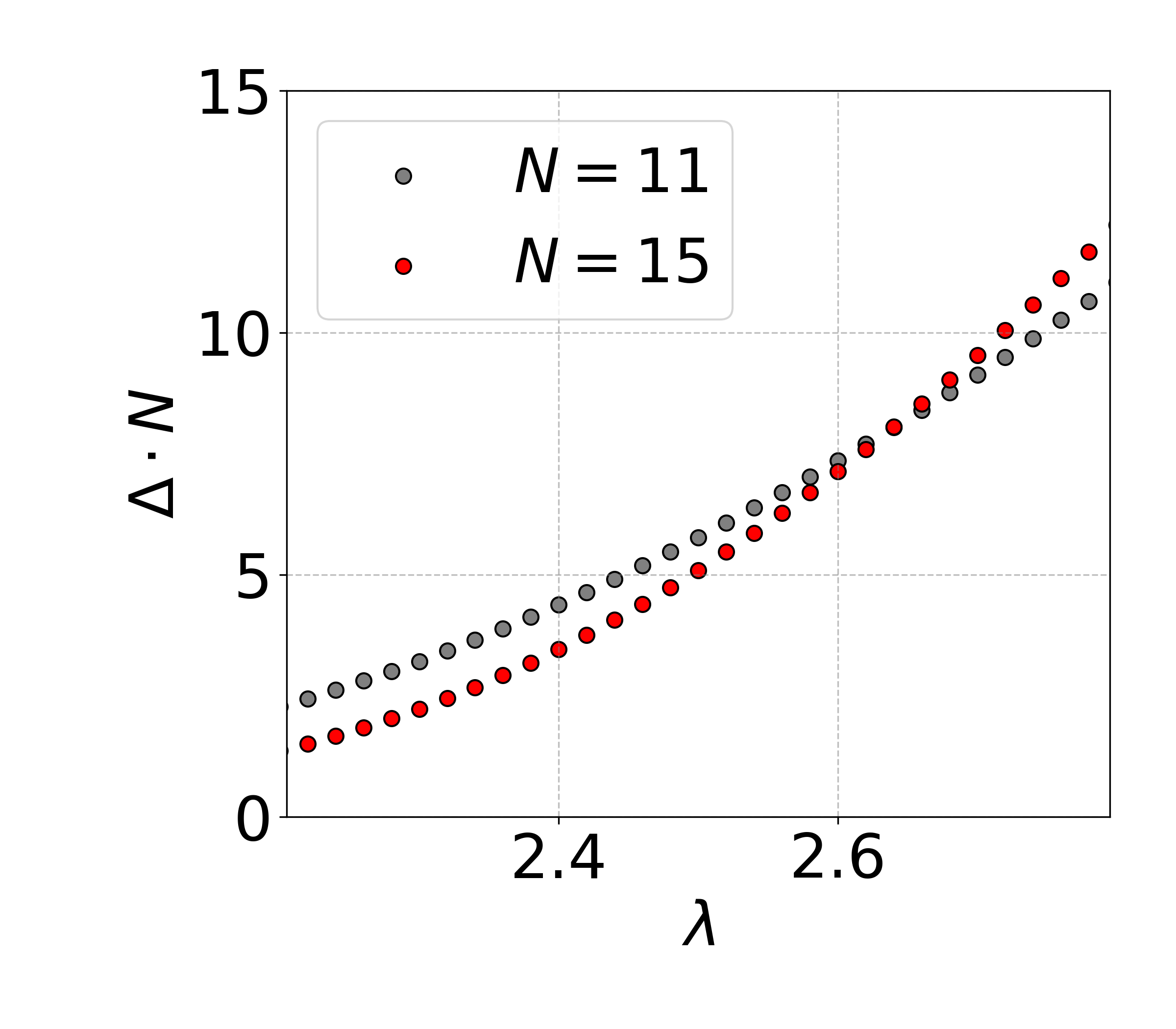}
\includegraphics[width=0.45\textwidth]{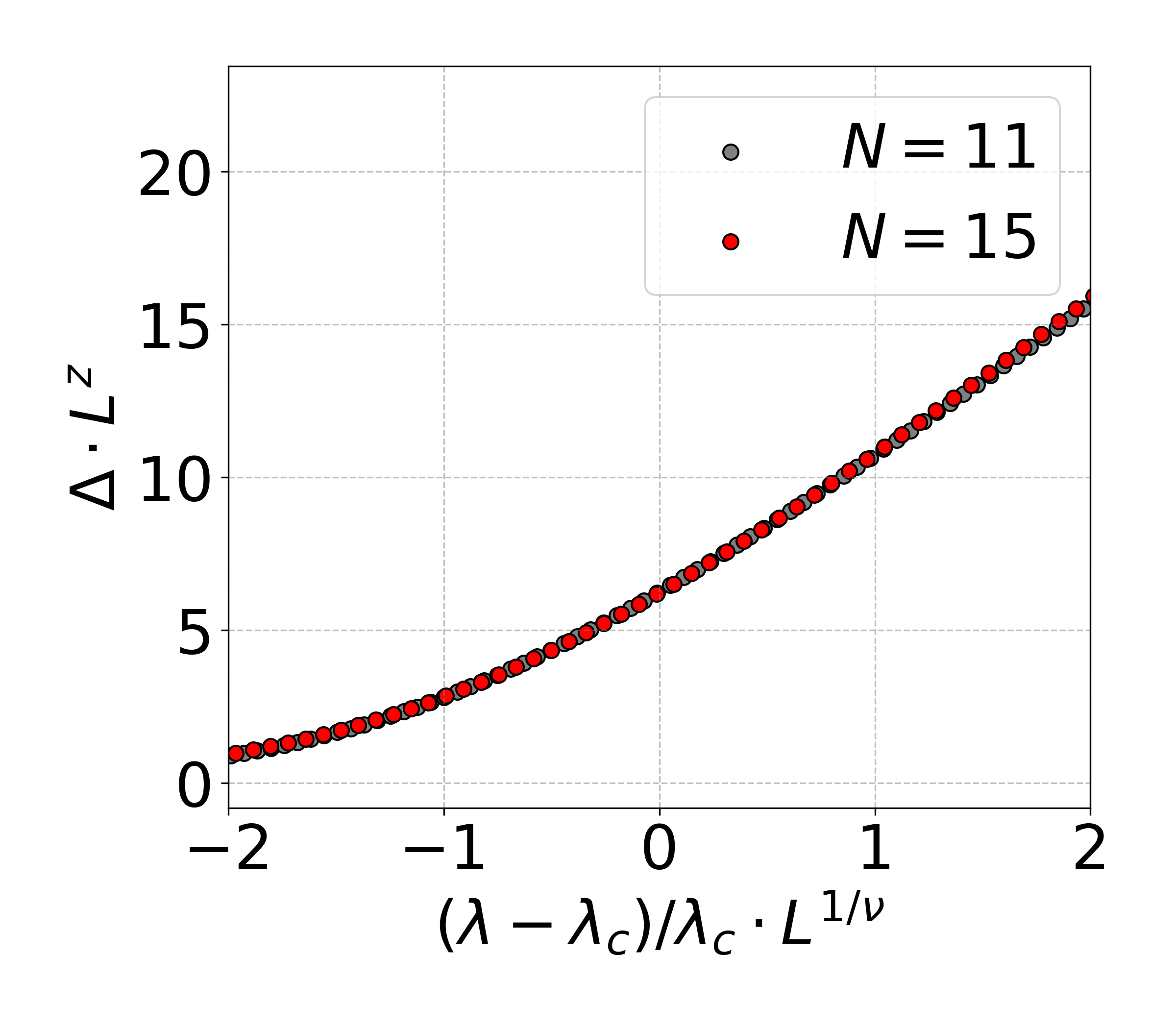}
\caption{The energy gap for the Sierpiński gasket before (left) and after rescaling (right). The scaling for $N = 11, 15$ with $\nu = 0.713(1)$ yields $\lambda_c = 2.703(4)$ and $z = 1.331(8)$.}
\label{fig:gasket_gap}
\end{figure}

Figure~\ref{fig:gasket_gap} presents the scaling collapse of the energy gap. Using the value of $\nu$ from the Binder cumulant analysis and treating $\lambda_c$ and $z$ as free parameters, we obtain $\lambda_c = 2.703(4)$ and $z = 1.331(8)$.

Finally, the susceptibility collapse gives $\lambda_c = 2.627(1)$, $\gamma = 1.671(1)$, and $\nu = 0.643(1)$.
Table \ref{Sierp_Gas_crit_exp} summarizes all FSS results for the two-size ($N = 11, 15$) collapses. The spread in $\lambda_c$ estimates from different observables ranges from 2.63 to 2.93.

Following Yi \cite{Yi2015}, we evaluate the scaling and effective dimension defined as $d_{\mathrm{scaling}}=(2\beta+\gamma)/\nu$ and $d_{\mathrm{eff}}=d_H+z$ respectively. Using the value of $\nu$ obtained from the Binder cumulant, a reasonable agreement is obtained $d_{\mathrm{scaling}}=3.19(2)$ and $d_{\mathrm{eff}}=2.92(1)$. This also shows, as expected, that standard errors resulting from fitting are smaller than errors resulting from finite-size sub-leading corrections.

\begin{table}[h]
\centering
\caption{Critical points and critical exponents obtained for TFIM
on Sierpi\'nski gasket lattice from finite-size scalling analysis of small systems.
Quantities used for FSS are in the columns, while critical parameters and critical exponents are in the rows.
The last column gives the width of the window in which FSS was performed.
The statistical errors referring to the last signiﬁcant digits are indicated in parentheses.}
\begin{tabular}{lcccccccc}
\toprule
\rowcolor{Gray}
            & $U$& $m$ &  $\Delta$ & $\chi$ & $\left| \widetilde{\lambda} L^\frac{1}{\nu} \right|$ \\
\midrule

15 - 11  \\
\midrule
 $\lambda_c$ & 2.7239(18) & 2.9304(57) & 2.7034(40)        &  2.6272(13) & $<\!1.0$ \\
 $\nu$       & 0.713(1) & 0.69(4) &  0.713(1)\footnote{To find $z$, we used $\nu$
obtained from scaling the Binder cumulant.}         & 0.643(1) & $<\!1.0$ \\
 $\beta $    & ---        & 0.3002(45) &   ---    & --- & $<\!1.0$ \\
 $\gamma$    & ---        & ---        &  ---                & 1.6706(10)          &$<\!1.0$ \\
 $z     $    & ---        & ---        &  1.331(8)            &  ---               & $<\!1.0$\\
\bottomrule
\end{tabular}

\label{Sierp_Gas_crit_exp}
\end{table}

\section{Numerical Renormalization Group method applied to TFIM on a 1D chain.}
\label{1D_TFIM_RG}

To validate our FSS results we used another method common in studies of critical phenomena – Numerical Renormalization Group (NRG). In this method the spins are blocked into non-interacting subsystems (blocks). Within each block, the ground state and first excited state are found by numerical diagonalization. In the next step of the renormalization procedure, each block is replaced by a single spin in a self-similar manner, and the Hamiltonian for the new system is obtained by rescaling the couplings.

This method has been successfully applied to our model example of 1D-TFIM by Jullien et al. \cite{Jullien1978}. We re-obtained their results for 1D-TFIM to prove the approach is valid and reproduces the correct critical parameters even for small cluster sizes. We then modified the method to calculate the critical properties of TFIM on the Sierpiński gasket.

In this approach, the spins are grouped into blocks of $N$ spins each, in which the Hamiltonian, Eq.~(\ref{Hamiltonian}), is diagonalized.
We label the ground and first excited states as $\ket{\tilde{0}}$, $\ket{\tilde{1}}$ respectively, with their energies labeled $E_0$, $E_1$. The diagonalized Hamiltonian has the form
\begin{equation}
    H=\sum_{j=1}^NE_j\ket{j}\bra{j}=E_0\ket{\tilde{0}}\bra{\tilde{0}}+E_1\ket{\tilde{1}}\bra{\tilde{1}}+\dots,
\end{equation}
which can be truncated to the two lowest energy states as follows
\begin{equation}\begin{split}
    \tilde{H}&=E_0\ket{\tilde{0}}\bra{\tilde{0}}+E_1\ket{\tilde{1}}\bra{\tilde{1}}\\&=-\frac{E_1-E_0}{2}\tilde\sigma^z+\frac{E_0+E_1}{2}\tilde{I},
\end{split}\end{equation}
where $\tilde\sigma^z$, $\tilde{I}$ are the "renormalized" spin-$z$ Pauli matrix and the identity in the two-dimensional, single-spin Hilbert space. We recognize this Hamiltonian describes a single spin coupled to a field, with the renormalized coupling
\begin{equation}
    \tilde{h}=\frac{E_1-E_0}{2}.
\end{equation}

To complete the renormalization procedure we include the coupling between adjacent blocks, $-J\sigma^x_i\sigma^x_j$. To this end we calculate the elements of  $\sigma^x_j$ between the $\ket{\tilde{0}}$, $\ket{\tilde{1}}$ states. The parity of energy eigenstates turns out to be crucial – since the terms in Hamiltonian (\ref{Hamiltonian}) cannot change the parity of an $N$-spin basis state, and the operator $\sigma^x_j$ flips a single spin, it becomes apparent that
\begin{equation}
    \bra{\tilde{0}}\sigma^x_j\ket{\tilde{0}}=\bra{\tilde{1}}\sigma^x_j\ket{\tilde{1}}=0.
\end{equation}
Since the block Hamiltonian is a real matrix, its basis vectors can be chosen to be real. $\sigma^x_j$ is real symmetric, so we can define
\begin{equation}\label{xi}
    \xi_j=\bra{\tilde{0}}\sigma^x_j\ket{\tilde{1}}=\bra{\tilde{1}}\sigma^x_j\ket{\tilde{0}}.
\end{equation}
The renormalized interaction term is thus of the form
\begin{equation}
    -(J\xi_i\xi_j)\widetilde{\sigma}_i^x\widetilde\sigma_j^x=-\widetilde{J}\widetilde{\sigma}_i^x\widetilde\sigma_j^x.
\end{equation}

\begin{figure}
    \centering
    \includegraphics[width=0.5\linewidth]{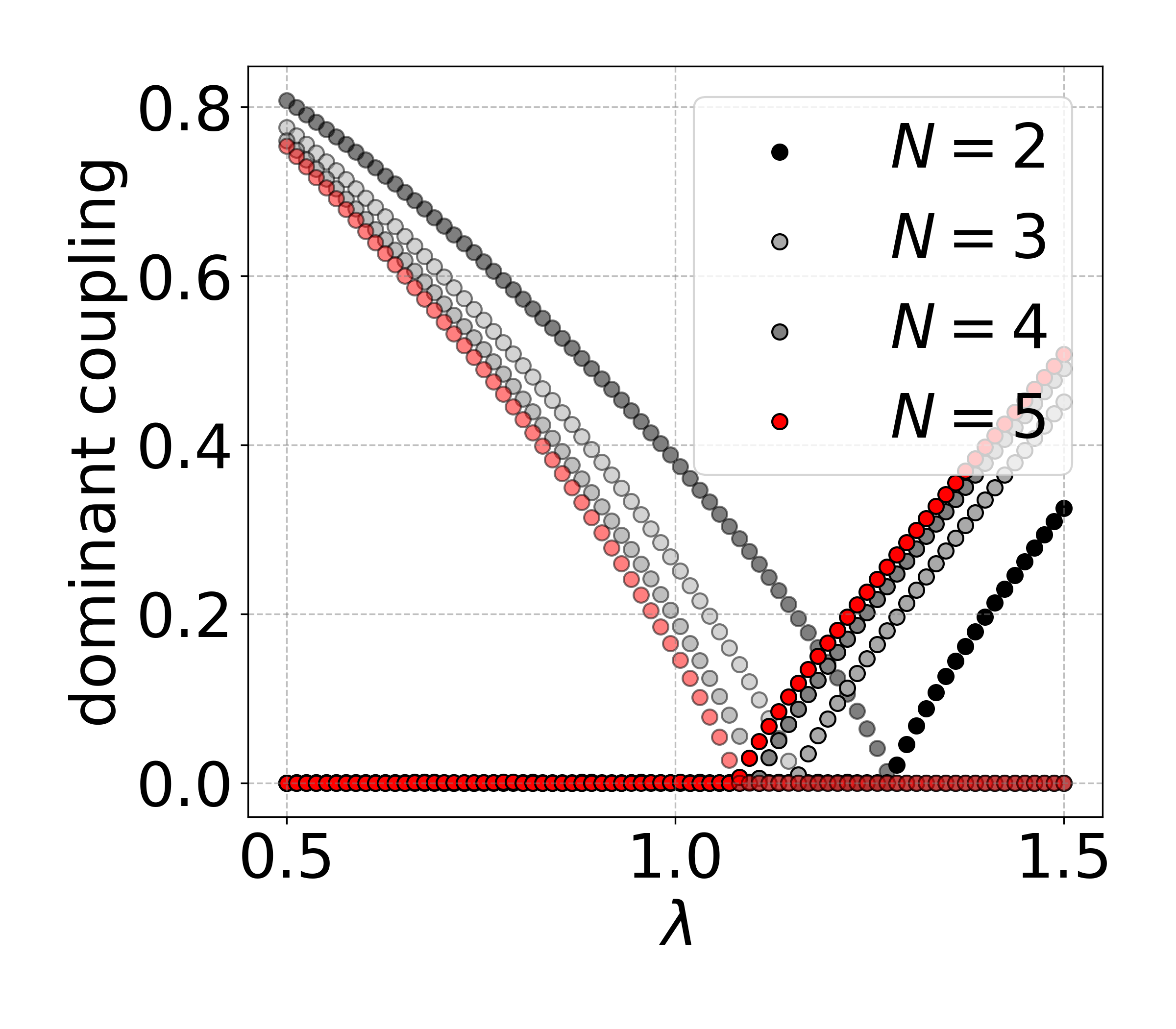}
    \caption{Renormalized couplings $J$ (to the left of the critical points, slightly transparent) and $h$ (to the right). We see that below critical coupling, spin-spin ordering dominates while magnetic field coupling is renormalized to $0$. Above the critical point the roles are reversed and the prevailing coupling is $h$, with $J\rightarrow 0$ with each step of the renormalization procedure.}
    \label{coupling renormalization}
\end{figure}

\begin{figure}
    \centering
    \includegraphics[width=0.4\linewidth]{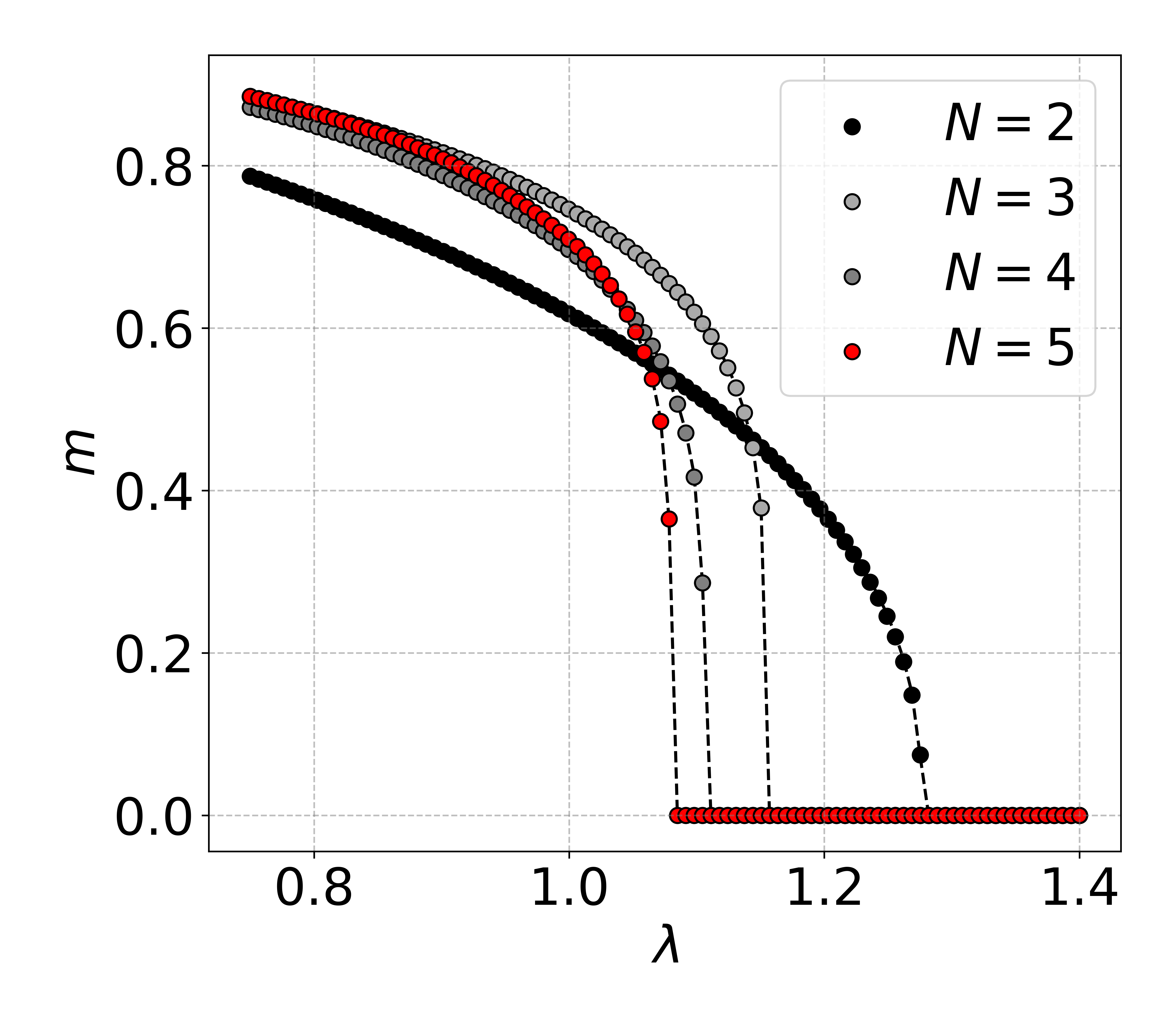}
    \caption{Renormalized magnetization for the 1D chain.}
    \label{1D chain renormalized magnetization}
\end{figure}

Having calculated the renormalized couplings we can write down the renormalization group equations. For the 1D-TFIM, the coupling between blocks is a single bond between the first and the last spin in the $N$-sized block. Due to symmetry the associated scaling factors $\xi_1$, $\xi_N$ must be equal and thus we can write the Hamiltonian at the $n$-th step of the renormalization procedure as
\begin{equation}
    H_n=-J_n\sum\widetilde\sigma^x_i\widetilde{\sigma}^x_{i+1}-h_n\sum\widetilde{\sigma}^z_i+C_n \widetilde{I}.
\end{equation}
With the couplings given by the following recurrence relations
\begin{equation}\begin{split}
    h_{n+1}&= \frac{E_{1,n}-E_{0,n}}{2},\\
    J_{n+1}&= \xi_{N,n}^2J_n,\\
    C_{n+1}&=N\,C_{n}+\frac{E_{0,n}+E_{1,n}}{2}.
\end{split}\end{equation}
The results of coupling renormalization can be seen in Fig. \ref{coupling renormalization}.
To calculate the magnetization it needs to be noted that the most representative member of the block is the spin in the middle. We can then calculate the renormalization of magnetization in a similar manner to the scaling factor of $\sigma^x$ in Eq.~(\ref{xi}) as follows
\begin{equation}
\begin{split}
m_{n+1}
    &= \left\langle \widetilde{\sigma}^x_{\mathrm{mid}} \right\rangle_{n+1} \\
    &= \xi_{\mathrm{mid}} \left\langle \widetilde{\sigma}^x_{\mathrm{mid}} \right\rangle_{n} \\
    &= \xi_{\mathrm{mid}}\, m_n .
\end{split}
\end{equation}
with $\xi_\text{mid.}$ equal to $\xi_{N/2}$ for $N$ even, $\xi_{(N+1)/2}$ for $N$ odd. The plots of renormalized magnetization for blocks of sizes $N=2,3,4,5$ can be seen in Fig. \ref{1D chain renormalized magnetization}.

 \section{Numerical Renormalization Group applied to the Sierpiński gasket}
 \label{Sierp_Gas_NRG}

In order to apply the NRG method to TFIM on the Sierpinski gasket we devised a suitable blocking procedure. We label the gasket at each step of construction as $T_1,\,T_2,\,\dots$ such that $T_1$ is a triangle, $T_2$ is three triangles and so on. A block $B_k$ is now defined as two $T_k$ gaskets with one mutual corner, and one other (inequivalent) corner removed from each. Since $T_k$ is made of $(3^k+3)/2$ points, then $B_k$ contains $N=3^k$ spins – we shall investigate only $N=3$ and $N=9$ spin blocks which can be efficiently diagonalized. The next step with $N=27$ would require us to diagonalize matrices of size $2^{27}\times 2^{27}$. However, having validated our method on the 1D chain, we can be reasonably sure of our results for $N=9$.

\begin{figure}
    \centering
    \includegraphics[width=0.4\linewidth]{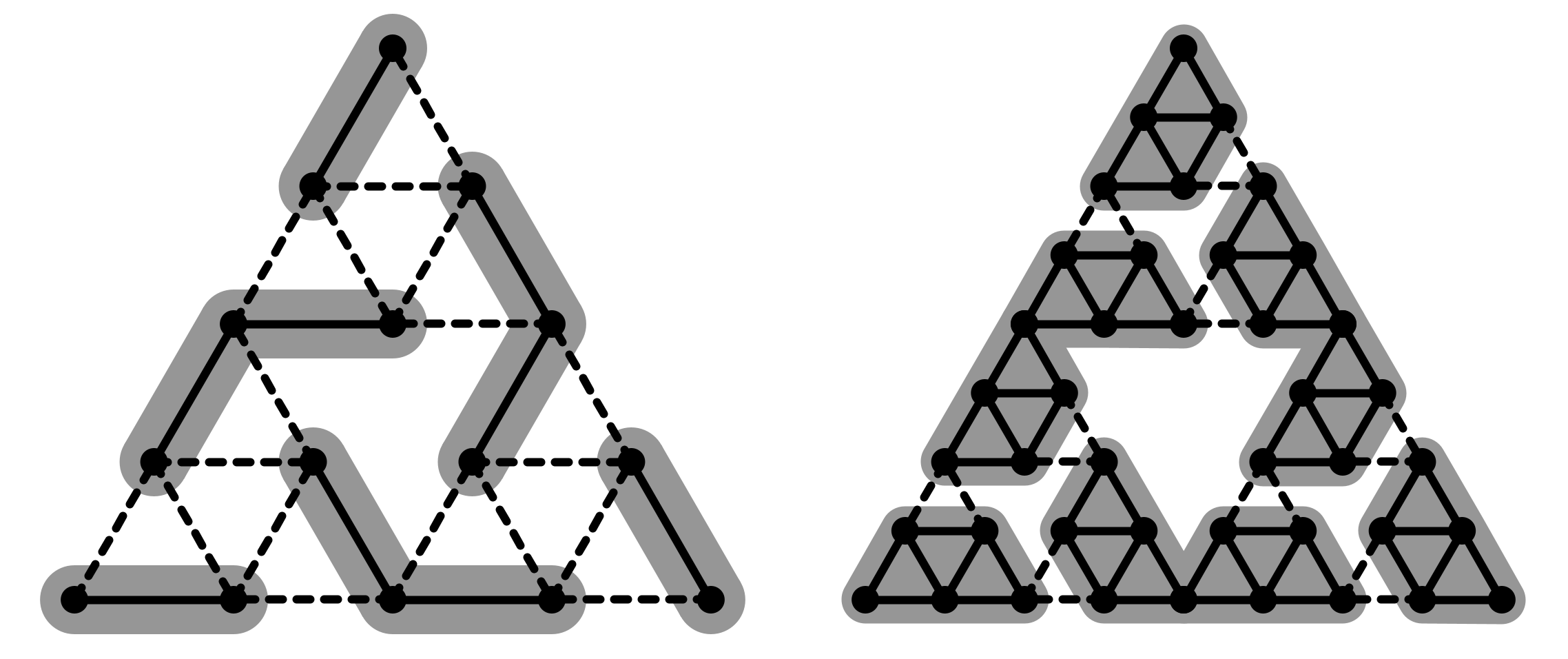}
    \caption{Blocking procedure for $N=3$ and $N=9$ spin blocks.}
    \label{sierpinski_spin_blocks}
\end{figure}

\begin{figure}
    \centering
    \includegraphics[width = 0.5\linewidth]{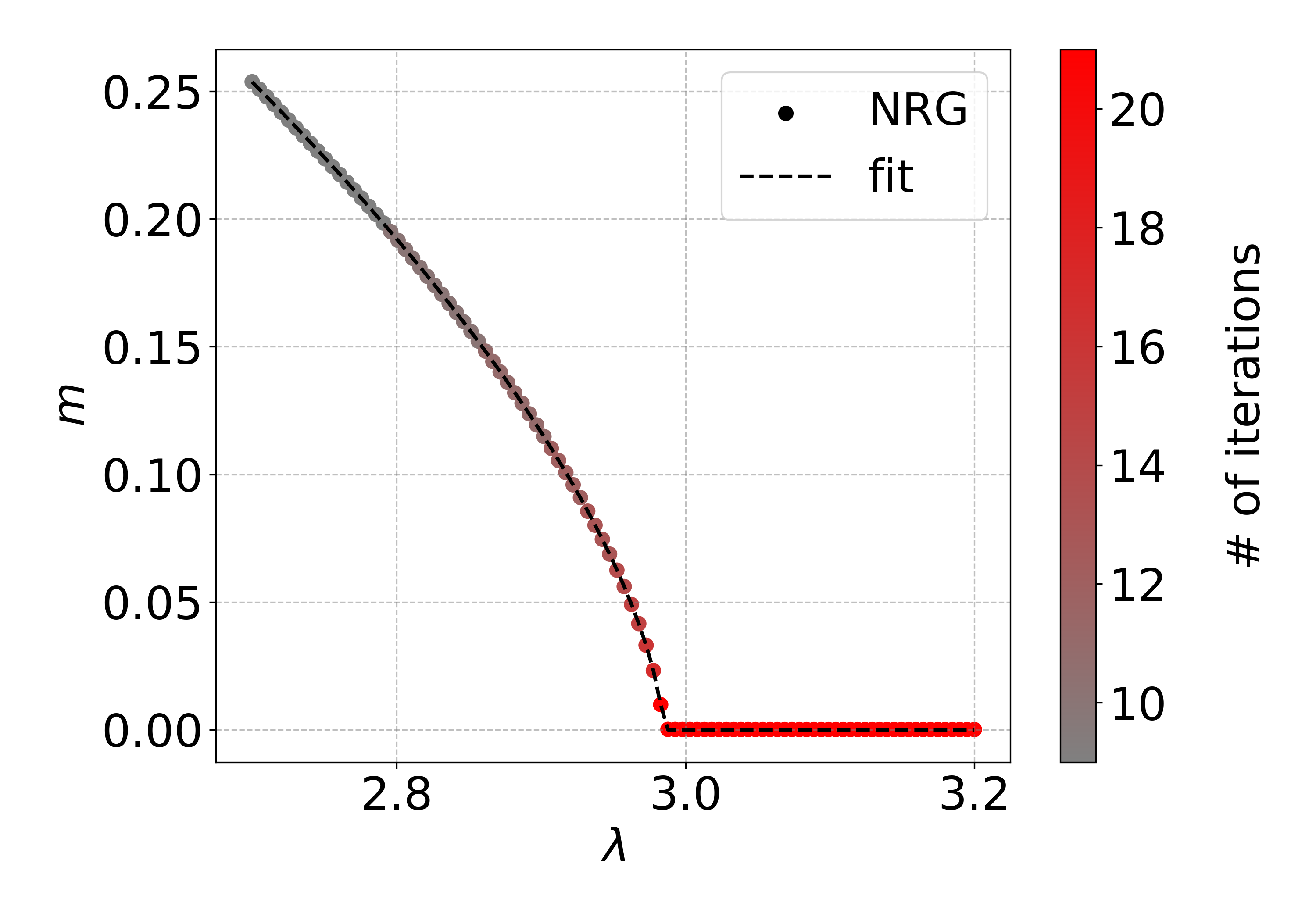}
    \caption{Renormalized magnetization for the Sierpiński gasket grouped into blocks of size $N=3$.}
    \label{gas_ren_mag_3}
\end{figure}

\begin{figure}
    \centering
    \includegraphics[width = 0.5\linewidth]{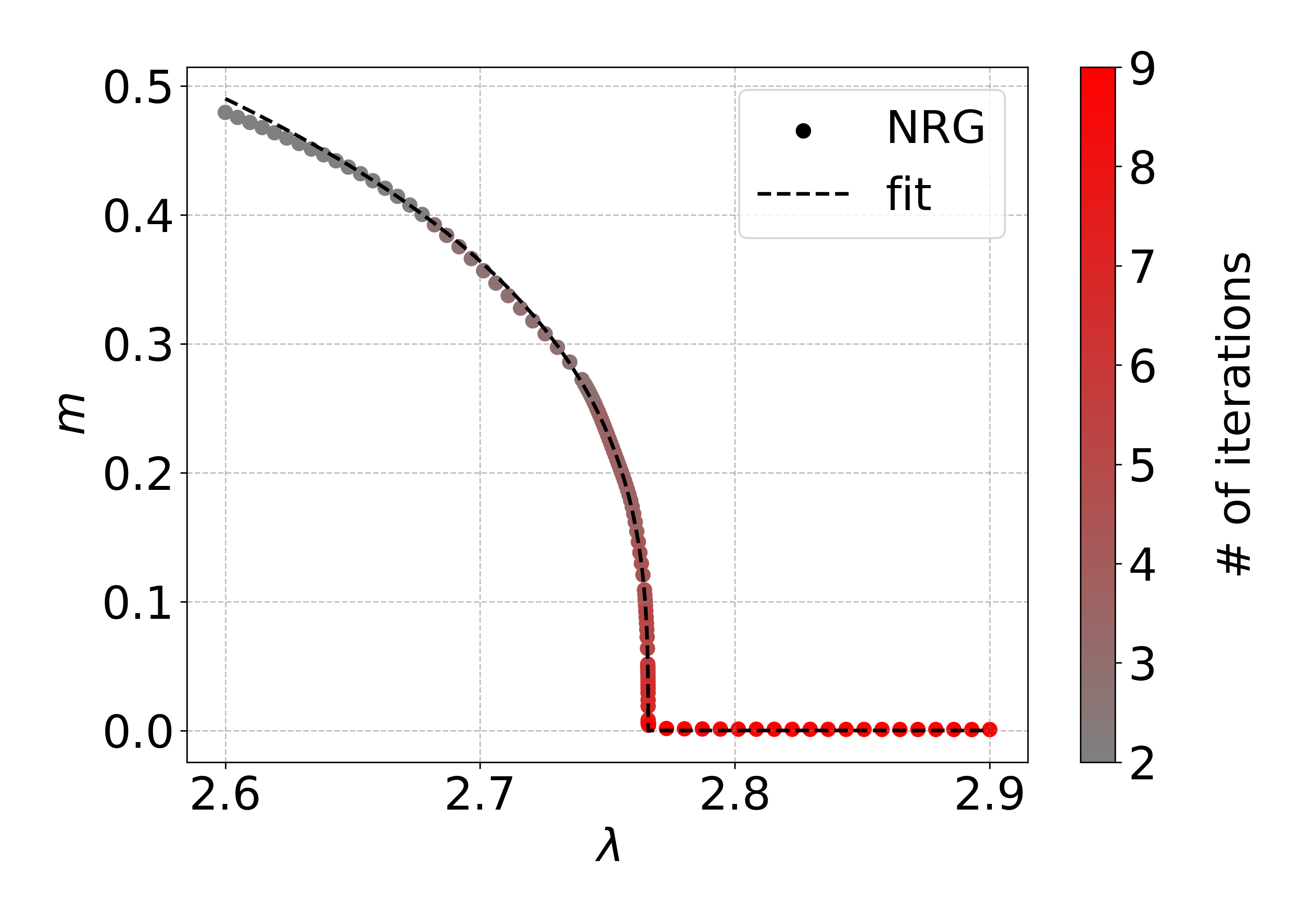}
    \caption{Renormalized magnetization for the Sierpiński gasket grouped into blocks of size $N=9$.}
    \label{gas_ren_mag_9}
\end{figure}

Since each renormalized bond is now composed of two original inter-block bonds, renormalization group equations need to be modified. Each new bond is composed of interactions between the outermost spin of the block with two of the inner spins (as can be seen in Fig. \ref{sierpinski_spin_blocks}). The renormalization group equations are thus
\begin{equation}\begin{split}
    h_{n+1}&= \frac{E_{1,n}-E_{0,n}}{2},\\
    J_{n+1}&= \xi_a(\xi_b+\xi_c)J_n,\\
    C_{n+1}&=N\,C_{n}+\frac{E_{0,n}+E_{1,n}}{2}.
\end{split}\end{equation}
The magnetization, resulting from above equations, is plotted in Fig. \ref{gas_ren_mag_3} and \ref{gas_ren_mag_9} for $N=3$ and $N=9$ respectively. The NRG algorithm is numerically unstable in the ferromagnetic ($\lambda < \lambda_c$) phase, which causes it to "fall off" the NRG flow after few iterations. Fortunately, the closer to critical point we get, the more resilient the algorithm becomes and the more iterations of the NRG algorithm we can afford (this is true for both 1D chain and the Sierpiński gasket).

\begin{figure}
    \centering
    \includegraphics[width = 0.5\linewidth]{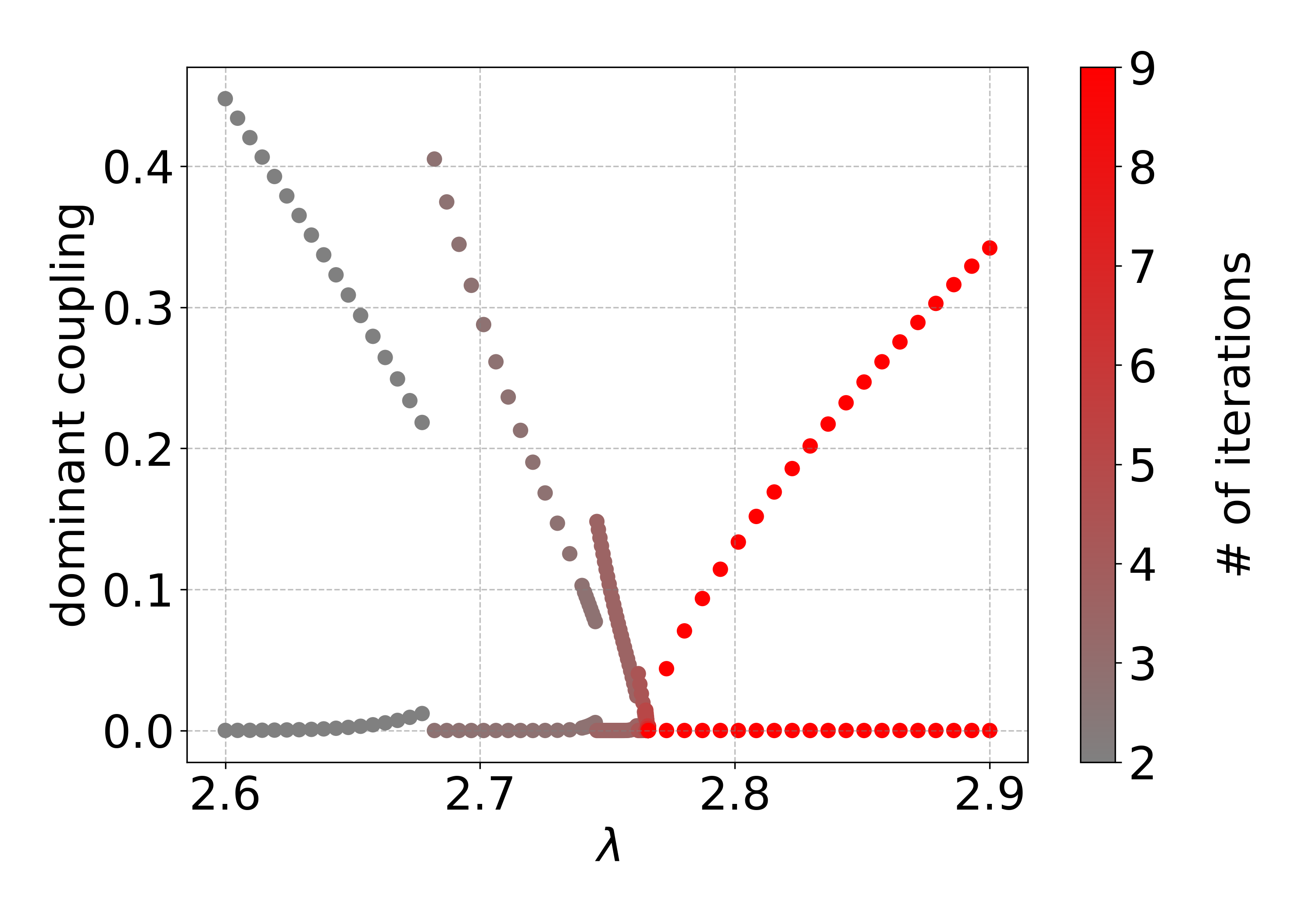}
    \caption{Renormalized couplings $J$ (to the left of the critical point) and $h$ (to the right) for blocks of size $N=9$. As in Fig. \ref{coupling renormalization}, $J$ dominates in the ferromagnetic phase. The discontinuities visible in this plot are a result of increasing number of NRG iterations as we approach $\lambda_c$ from below (in Fig. \ref{coupling renormalization} the number of iterations was fixed).}
    \label{gas_ren_couplings}
\end{figure}
The critical coupling strength and $\beta$ exponent can be found by fitting a power law $m\propto (\lambda-\lambda_c)^\beta$ to the magnetization values in the vicinity of
$\lambda_c$. Table \ref{NRG results} collects the fitted parameters for both 1D chain and the Sierpiński gasket. As can be seen for the 1D chain, the results get closer to the exact values of $\lambda_c=1$ and $\beta = 0.125$ the bigger $N$ is, which is expected. For $N=5$ blocking of the 1D chain the ratio of inter-block bonds to intra-block ones is $0.5$, while for the $N=9$ blocks on the gasket, the ratio is $9/14\approx0.64$, which is comparable, so the results for the gasket can be trusted (the method works better when this ratio is as small as possible).

\begin{table}
\centering
\caption{
Critical coupling and $\beta$ exponent obtained using the NRG method for 1D chain and the Sierpiński gasket. }\label{NRG results}
\begin{tabular}{lcccc}
\toprule
\rowcolor{Gray} \cellcolor{Gray} 1D chain & $N\!=\!2$ & $N\!=\!3$ &$N\!=\!4$ & $N\!=\!5$\\
\midrule
$\lambda_c$ & 1.276 & 1.154 & 1.106& 1.080\\
 $\beta $    &   0.393    &0.182 &0.191 &0.162 \\
\midrule
\rowcolor{Gray} Sierpiński gasket & $N\!=\!3$\! &&  $\!N=\!9$\!   &\\
\midrule
 $\lambda_c$&  2.985 && 2.766   \\
 $\beta $   &   0.641 && 0.316 \\
\bottomrule
\end{tabular}
\label{MAO_MGO_Model4}
\end{table}

\section{Conclusion}
\label{Summ}

We demonstrated that successful curve collapses, being the heart of the finite-size scaling (FSS) method, can be obtained even for very small system sizes for fractal lattices. In our case it was $ N = 11, 15$ for the Sierpiński gasket. Obtained critical field $\lambda_c \approx 2.63 - 2.93$ and exponents $ \nu \approx 0.64 - 0.71, \beta \approx 0.30, \gamma \approx 1.67, z \approx 1.33$ need an independent verification, to show that small system sizes can indeed give a useful rough approximation to the listed parameters. We provided it by means of the numerical renormalization group method. Even though it also works with finite cluster sizes, it is burdened only with Hilbert space truncation errors and beside it, deals practically with the thermodynamic limit. Thus it should be regarded as an entirely independent method. It gave $\lambda_c = 2.766$ and $\beta = 0.316$, supporting FSS and usefulness of small systems, even for fractal lattices. Both methods were also checked on the one-dimensional transverse-field Ising model.

Our results, when compared to those of Yi \cite{Yi2015} and Krcmar et al. \cite{Krcmar2018} give a significantly different critical field $\lambda_c$ (their value is $\lambda_c = 1.865(2)$). This confirms, that Yi most likely used the same lattice geometry as Krcmar et al. A significant increase in $\lambda_c$ for our lattice is an expected result, since the coordination number is increased from 3 to 4.
Regarding critical exponents, our values of $z$ and $\nu$ are very similar to those of Yi \cite{Yi2015}, especially taking into account approximate character of our estimates. Exponents $\beta$ and $\gamma$ are more deviated, which suggests that whether two versions of the Sierpiński gasket belong to the same universality class in not obvious, but our results do not settle it decisively. Comparison of the scaling and effective dimensions supports consistency of the determined critical exponents.


\begin{references}
\bibitem{Gefen1981} Y.~Gefen, A.~Aharony, B.~B.~Mandelbrot, and S.~Kirkpatrick, Phys.\ Rev.\ Lett.\ \textbf{47}, 1771 (1981).
\bibitem{Gefen1984} Y.~Gefen, A.~Aharony, and B.~B.~Mandelbrot, J.\ Phys.\ A \textbf{17}, 1277 (1984).
\bibitem{Jullien1978} R.~Jullien, P.~Pfeuty, J.N.~ Fields, and S.~Doniach, Phys. Rev. B \textbf{18} 3568 (1978).
\bibitem{Tomczak1994} P.~Tomczak, W.~Je\.zewski, Physica A \textbf{209)}, 275 (1994).
\bibitem{Kovacs2010} I.~A.~Kov\'acs and F.~Igl\'oi, Phys.\ Rev.\ B \textbf{82}, 054437 (2010).
\bibitem{Yi2015} Hangmo~Yi,  Phys.\ Rev.\ E \textbf{91}, 012118 (2015).
\bibitem{Krcmar2018} R.~Krcmar, J.~Genzor, Y.~Lee,  H.~\v{C}en\v{c}arikov\'a,
T.~Nishino, and A.~Gendiar, Phys. Rev. E \textbf{98}, 062114 (2018).
\end{references}
\end{document}